# Schooling and Labor Market Consequences of School Construction in Indonesia: Comment

David Roodman[*]

**Abstract**

Duflo (2001) exploits a 1970s schooling expansion in Indonesia to estimate impacts on schooling and labor outcomes, as well as the returns to schooling. I correct data errors, adjust for potential sources of bias, follow up later in life, and check sensitivity to two specification choices that are not explained in the original text and that are applied to some regressions and not others. Headline estimates from the original, 1995 follow-up are generally not robust to the revisions. Returns to schooling are weakly identified. But certain "fingerprints" from Duflo (2001) specification checks—impacts appearing only in the birth cohorts or at the grade levels where expected—recognizably persist. E.g., impact estimates from kink-based specifications that focus on timing by birth cohort are consistently positive, if not always at conventional significance levels. The best explanation, though weakened, is causation from the schooling expansion to individual schooling attainment to wages.

[*] Open Philanthropy, 182 Howard St., #225, San Francisco, CA 94105 (e-mail: david.roodman@openphilanthropy.org). Thanks to Esther Duflo for data, code, and comments; to Daniel Feenberg at the National Bureau of Economic Research for running and troubleshooting scripts; and to Mikey Jarrell, Nathan Lazarus, Jonathan Morduch, Otis Reid, Justin Sandefur, and anonymous referees for comments. Estimates and inferences are performed with boottest (Roodman et al. 2019) cic (Melly and Santangelo 2015), and reghdfejl (Roodman 2025). Tables and figures were made with blindschemes (Bischof 2017), coefplot (Jann 2014), esttab (Jann 2007), and palettes (Jann 2023).

## Introduction

In 1973, the government of Indonesia dedicated part of an oil windfall to a project to erect thousands of schoolhouses across the nation's far-flung archipelagic domain. The Instruksi Presiden Sekolah Dasar program (Inpres SD) became one of the largest schooling expansions ever. In its first six years alone, it doubled the country's stock of primary schools (Duflo 2001). Following up on affected men in early adulthood, Duflo (2001, p. 812) concludes that Inpres SD "was effective in increasing both education and wages."

Duflo (2001) is influential for several reasons. Typifying its moment in intellectual history, it exploits strategies for causal inference such as difference-in-differences (DID) and natural experiment–based instrumentation (Angrist and Krueger 1999). It innovates, in part by applying such methods to a developing-world setting. The natural experiment it brings to light dwarfs most in the education literature. Threats to identification are confronted: a placebo check in the pre-treatment period returns a null result; the instruments pass a validity test; other potential sources of bias are named and addressed. Graphical analysis makes the case that the schooling shock leaves distinctive fingerprints in the data, in particular that it manifests with the expected timing and at the expected grade levels. Few studies so credibly claim to identify the impacts of schooling at scale.

Precisely because of its continuing importance, I revisit the Duflo (2001) quasi-experiment. The passage of time has brought improvements in econometric methods. It has brought more data too, as the subjects of the quasi-experiment have transited through their labor careers, leaving traces in regular natural surveys. And any fresh look may add insight.

The reanalysis proceeds in two steps. First, the data set is reconstructed from primary sources and the original analysis is reimplemented in code. Comparing the original and new versions surfaces bugs in both and builds confidence in the (debugged) new version.

Second, the Duflo (2001) specifications are modified in several ways to check robustness and improve inference. A rational reader will put some weight on the possibility that in assembling these modifications, I have "reverse *p*-hacked," trying many and selectively reporting, consciously or not, the most provocative results. Like Duflo (2001), this reanalysis was not pre-registered. To combat the threat and the perception of reverse *p*-hacking, the modifications are chosen to minimize consumption of "researcher degrees of freedom" (Simmons, Nelson, and Simonsohn 2011). The study population is not modified, nor the sample by, say, removal of outliers. No new controls are added. The original estimators are largely hewed to. Some modifications are borrowed from my earlier work rather than fabricated anew for this project.

The main changes resulting from both steps are:

1. Correcting data entry errors in baseline variables, which affect about 10% of observations.
2. Clustering standard errors, by the geographic unit of treatment.
3. Incorporating later data on the same subjects, as in the Easterly, Levine, and Roodman (2004) comment on Burnside and Dollar (2000). This increases the power to detect impacts on schooling attainment and adds information on labor outcomes in the prime working years.
4. Applying now-standard methods to test for and accommodate weak identification in instrumental variables (IV) estimation, in particular the Kleibergen-Paap (2006) measure of identification strength and a bootstrapped Anderson-Rubin test to construct confidence intervals.
5. Using a two-piece linear spline model ("kink model") to formally test an informal characterization in the original text about the presence of a trend break with timing attributable to Inpres SD.
6. Copying specification choices from some regressions to others within the original paper, as in the Roodman (2018) comment on Bleakley (2007)—choices that are in this case essentially undiscussed in the original. The choices are two: whether to incorporate survey weights and whether to include a particular vector of controls.

The first four items are relatively uncontroversial. Clustering by geographic treatment unit when performing DID on microdata has been standard since Bertrand, Duflo, and Mullainathan (2004). Duflo (2004a) also follows up later on the same population. Since 2001, the literature on weak identification in IV estimation has shifted norms among practitioners.

The last two items are more debatable, for each makes a trade-off of unknown degree between bias and power. Incorporating survey weights reduces the risk of bias from endogenous survey sampling, but tends to reduce precision. Controlling for pre-treatment trends through the kink model should reduce bias from non-parallel trends, but also consumes power since it asks not merely whether a post-treatment trend was positive, but whether the trend increased. It is hard to know whether dropping a set of controls raises or lowers the exogenous fraction of identifying variation. Because these changes are more matters of judgment, they are performed and reported in all combinations. If they were introduced cumulatively in a particular order, the choice of that order would consume researcher degrees of freedom, justifiably raising a question in the reader's mind about whether it was favored in order to create a certain impression.

Duflo (2001)'s headline ordinary least squares (OLS) findings—0.12–0.19 more years of schooling per new school per 1,000 children and 1.5–2.7% higher wages—both fade with data corrections, clustering, and reversal of the two unexplained specification choices. The point estimates fall, while, for schooling, the corresponding "placebo" impact rises. The impact on schooling attainment is ambiguous enough that instrumental variables estimates of the returns to schooling often suffer from weak identification, while weak-identification-robust inference sometimes produces infinite confidence intervals. The story does not fundamentally change when adding data from later follow-ups.

Yet, especially in those larger samples, the evidence from the Duflo (2001) fingerprint checks cuts against the null of zero impacts. In the kink-based regressions, the point estimates are almost always positive: they are almost always positive for educational outcomes (total years of schooling and completion of primary school) and for labor market outcomes (formal-sector employment and wages); they are almost always positive with or without survey weights, and with or without a questioned set of controls. True, many of the new estimates are not significant at conventional levels, especially when the questioned control set is dropped. But there is statistically strong evidence that the long-term rise in the fraction of boys completing primary school accelerated in places that got more new schools per child, relative to areas that got fewer—and with timing attributable to Inpres SD. A similar kink is found for wages, and with similar strength, at least when the questioned controls are included.

For these reasons, the hypothesis that Inpres SD increased schooling and thereby wages remains the best explanation for the evidence at hand. But the theory does not earn the same credence as in the original study. And the returns to Inpres-stimulated schooling are hard to measure with any precision.

Section 1 of this paper reviews the methods of Duflo (2001). Section 2 describes and motivates the changes to data and method. Section 3 replicates key Duflo (2001) results and documents the effects of changes in the original, 1995 follow-up. Section 4 moves to later follow-ups. Section 5 brings the same revisions to the analysis of schooling impact by grade level. Section 6 summarizes results from changes-in-changes (CIC), an estimator that is designed to accommodate non-parallel trends rather than model and control for them. Section 7 concludes.

## 1 The Duflo (2001) specifications

Authority to carry out the Inpres SD program flowed from annual presidential instructions, appendices of which set construction targets for every *kabupaten* (regency) and *kotamadya* (municipality), which are Indonesia's second-level administrative units, and which I call "regencies" for short. Duflo (2001)'s treatment intensity indicator is the number of schools planned for construction in a regency between 1973/74 and 1978/79, per 1,000 children aged 5–14. Duflo (2001, note 1) cites a government finding that in this period, actual construction closely matched planned. Using data from the nationally representative Intercensal Population Survey (SUPAS) of 1995, the study examines how boys from less- and more-treated regencies fared in early adulthood: how many years they ultimately stayed in school; whether they held wage employment; and, if so, how much they earned per hour and month.

Duflo (2001) works in the difference-in-differences framework. Since the outcome data come from a single, 1995 cross-section, the study arrays observations into a two-dimensional structure by grouping on year of birth and place of birth, as in the Acemoglu and Angrist (2000) study set in the U.S. The oldest subjects are the pre-treatment cohorts since Inpres SD was launched too late to directly affect them. Analytical time therefore runs from older to younger people. The included cohorts range in age from 2 to 24 as of 1974.

To illustrate ideas, Duflo (2001) begins with classical 2×2 DID. It is performed by fitting

$$Y_{ijt} = D_j T_t \delta + \nu_j + \eta_t + \epsilon_{ijt} \tag{1}$$

where $Y_{ijt}$ is an outcome for individual $i$ in treatment group $j$ in period $t$; $j = 0, 1$ for the control and treatment group; $t = 0, 1$ for the pre and post-treatment periods; $D_j$ is a dummy for being from a high-treatment regency; $T_t$ is a dummy for the younger, treated cohorts; $\delta$ is the impact parameter; $\nu_j$ and $\eta_t$ are treatment group and cohort fixed effects; and $\epsilon_{ijt}$ is an exogenous, mean-zero error.

Next, Duflo (2001) elaborates beyond the 2×2 grid. Each yearly birth cohort gets its own time index and each of the approximately 290 regencies in Indonesia as of 1995 becomes its own treatment group. Treatment is taken as continuous, as planned schools per 1,000 children. These extensions lead to the two-way fixed effects (TWFE) model,

$$Y_{ijt} = \left(D_j \widetilde{\mathbf{T}}_t\right)' \boldsymbol{\delta} + \nu_j + \eta_t + \epsilon_{ijt} \tag{2}$$

where $t$ and $j$ index birth cohorts and regencies and $D_j$ is now treatment intensity in regency $j$. $\widetilde{\mathbf{T}}_t$ is a vector of variables that depend only on time, i.e., on age in 1974. Duflo (2001) defines $\widetilde{\mathbf{T}}_t$ in several ways. In its least flexible form, $\widetilde{\mathbf{T}}_t$ is the $T_t$ defined earlier, a single dummy for the younger cohorts. In this case, men aged 2–6 in 1974 form the "after" cohorts and those aged 12–17 the "before" cohorts while the rest are dropped. (The slight asymmetry in these definitions, with one age range wider than the other, is not discussed in the original text.) In its most flexible form, $\widetilde{\mathbf{T}}_t$ is a set of yearly cohort dummies, and all cohorts aged 2–24 constitute the sample. The estimate $\hat{\boldsymbol{\delta}}$ is then an event study, i.e., a vector of cohort-specific linear associations between treatment and outcome. An in-between option that Duflo (2001) sometimes favors makes $\widetilde{\mathbf{T}}_t$ a set of yearly cohort dummies, except that the pre-treatment cohorts aged 12–24 in 1974 are given one dummy.

Alongside the treatment terms and fixed-effect dummies, most Duflo (2001) TWFE specifications include controls. All are products of baseline variables and cohort dummies, to allow the variables cohort-specific associations with the outcome. The augmented specifications read

$$Y_{ijt} = \left(D_j \widetilde{\mathbf{T}}_t\right)' \boldsymbol{\delta} + (\mathbf{C}_j \otimes \mathbf{T}_t)' \boldsymbol{\gamma} + \nu_j + \eta_t + \epsilon_{ijt} \tag{3}$$

$\mathbf{T}_t$ is a full set of cohort dummies. In successive variants, $\mathbf{C}_j$ grows from one to three variables. It starts with the number of children in a regency in 1971 aged 5–14, which is the denominator of the treatment indicator. It gains the fraction of the population enrolled in primary school, to reduce bias from mean reversion. Then it gains regency-level spending on a separate water and sanitation program, which might confound the effects of school construction. I will call these variants the minimal, intermediate, and full control sets.

A final step in the Duflo (2001) empirics is to perform two-stage least squares (2SLS) to estimate the returns to schooling. The 2SLS specifications use (3) for the first stage, setting $Y_{ijt}$ to years of schooling ($S_{ijt}$).[1] They adapt (3) for the second stage, replacing $D_j \widetilde{\mathbf{T}}_t$ with the instrumented variable, $S_{ijt}$, and setting $Y_{ijt}$ to either a dichotomous indicator of wage sector employment or to log monthly or hourly wages. Wage regressions are normally restricted to participants in the wage sector, although in a robustness test, wages of non-wage workers are imputed. $\widetilde{\mathbf{T}}_t$ takes two of the forms described earlier: binary, and having one dummy for ages 12–24 and another for each yearly age from 11 down to 2. I call these the "instrument by young/old" and "instruments by birth year" variants.

One threat to identification in DID is the presence of non-parallel trends—that is, distinctive trends in outcomes within the various treatment arms that are separate from the causal story of interest. They may be visible before treatment and persist after. A standard specification check is therefore for relative trends before treatment, or "pre-trends." Duflo (2001, Tables 3 & 4) checks for pre-trends through a placebo or "control" experiment in which the pre- and post-treatment periods are shifted to lie fully within the true pre-treatment era. Those aged 18–24 in 1974 become the pre-treatment group and the actual pre-treatment group, aged 12–17, becomes the post-treatment group for this test.

Alongside the regression results, Duflo (2001) disaggregates the specifications along one dimension or another and plots the results to check for fingerprints of the intervention. By surfacing patterns that would unlikely occur by chance, these increase confidence in the main results. One instance, already mentioned, is the event study, in which the treatment variable is interacted with a full set of birth cohort dummies, and the coefficients plotted. Here, it appears that the intermediate control set is used.[2] The other fingerprint check decomposes schooling attainment into a set of survival dummies: for whether a person attended school for at least one year, or at least two, etc. These dummies become dependent variables in the TWFE specification (2), the one without extra controls. The resulting plot of "difference in differences in CDF [cumulative distribution function]" checks whether the primary schooling expansion indeed manifested through greater continuation through primary school.[3]

[1] A more precise description of $S_{ijt}$ is "highest year of school attended." The two concepts can diverge. Suppose the construction of a new primary school causes a child to switch to it from a farther-away middle school. And suppose that the child exits schooling upon graduation from that primary school. Later in life that child could report a lower "highest year of school attended" without any reduction in years of schooling.

## 2 Comments

I comment on seven aspects of the Duflo (2001) regressions: data transcription errors, clustering, the possibility of heterogeneous treatment effects, a vector of controls not motivated in the text, weak identification, observation weighting, and the handling of potentially non-parallel trends. The comment on effect heterogeneity leads to a reassuring diagnostic check. The rest lead to revisions in data and method.[4]

### 2.1 Data corrections

Duflo (2001) gathers early-1970s regency-level variables from government publications. For example, reports on the 1971 census provide total population by regency and population aged 5–14. Both are denominators for other variables.

I reconstruct all the 1970s variables but the one on water and sanitation spending, which is used in only some specifications and does not affect results much.[5] I find some discrepancies between the Duflo (2001) figures and official sources (BPS 1972, 1974; Bappenas 1973–78). The transcription errors affect the treatment indicator for about 10% of regencies. For example, probably because regencies are listed in different orders in different publications, a few pairs or quadruplets have their numbers rearranged. The original and corrected versions of the treatment intensity indicator $D_j$ are correlated 0.81 across regencies.[6] Duflo (2001, Table 2) confirms that Inpres SD targeted areas with many young children not enrolled in primary school, by reporting a regression of the log number of schools planned for each regency on the log number of children aged 5–14 and the log primary school non-enrolment rate; with the corrections, the $R^2$ rises from 0.78 to 0.83.

[2] This is inferred by comparing to Duflo (1999, chapter 1, Table A1).

[3] Duflo (2001, p. 804) defines the dependent variables as the dummies $\mathbf{1}\{S_{ijt} \leq m\}$, i.e., whether individual $i$ completed $m$ years of schooling or less. In fact, the dummies appear to be $\mathbf{1}\{S_{ijt} > m\}$, which merely flips the signs on the estimates.

[4] Previous versions of this paper mention all of these issues. However, one, the vector of controls not motivated in text, was touched on only briefly. The first commit of the code to GitHub, in October 2022, loops over control sets: no controls, and the minimal, intermediate, and full control sets. But I initially dropped most discussion of no-controls results because they did not seem important for kink-based inference, which I most emphasized. On reflection, it seemed appropriate to highlight the influence of a specification choice not motivated in the original text.

[5] Page scans of primary sources and an annotated Excel tabulation are at github.com/droodman/Duflo-2001/tree/main/Printed%20sources.

[6] A particular complication affects baseline school attendance. Duflo (2001) appears to take the numerator, people aged 5 and older attending school, from BPS (1974a), and the denominator, total population, from a different publication, such as BPS (1972). The figures for Irian Jaya (Papua) are much lower across the board in BPS (1974a), perhaps because school attendance was only queried for a subpopulation. For consistency, I therefore use the denominator that accompanies the numerator in BPS (1974a), namely, total population aged 5 and above.

In response to an earlier version of this paper, Esther Duflo, Mikey Jarrell, and Nathan Lazarus endorsed the corrections.

### 2.2 Clustering

The variance estimates in Duflo (2001) are classical: they are not adjusted for the possibilities of heteroskedasticity and intra-regency and intertemporal dependence. In the other chapters of the Duflo (1999) thesis from which Duflo (2001) is derived, standard errors are clustered. And Bertrand, Duflo, and Mullainathan (2004) demonstrates the value of the clustering correction (Liang and Zeger 1986, Arellano 1987) when performing TWFE on microdata with treatment assigned by geography.

### 2.3 Heterogeneous treatment effects

Recent scholarship highlights the counterintuitive way that effect heterogeneity can bias TWFE estimates of mean effects (de Chaisemartin and D'Haultfœuille 2020; Goodman-Bacon 2021). This methodological critique does not apply with force to Duflo (2001). The scope for bias is greatest when treatment is staggered across units, unlike here. Moreover, Jakiela (2021)'s check for heterogeneity produces little evidence of such: see appendix A.

### 2.4 Controls based on the number of children in the regency of birth

As described in section 1, most TWFE regressions in Duflo (2001) include in $\mathbf{C}_j$ the number of children aged 5–14 in the regency of birth as of 1971. The variable is interacted with the cohort dummies to control for cohort-specific associations between population and the outcomes.

This specification choice warrants three comments. First, it is unusual: the norm is to enter a population variable in logarithms.

Second, the controls are not always included. While they enter the main TWFE regressions, they are dropped when the dependent variables are dummies for having completed at least a certain number of years of schooling (Duflo 2001, Figure 2).

Third, the controls receive little mention and even less explanation; and the bit of explanation suggests that the controls are included with an eye toward their effect on the results. The Duflo (1999) thesis chapter does not mention the controls except in footnotes to tables, and even describes one set of results that evidently include them as "without controls."[7] The Duflo (2000) working paper version mentions the controls in text, much as in the published paper. And unlike in the publication, the working paper provides some motivation, in an endnote: "The number of schools per child was negatively correlated with the number of children, and I want to abstract from any time-varying effect of regional population. It turns out to be important in the wage equation."

[7] Duflo (1999, p. 34). The reference is to the chapter's Table 5, whose notes do *not* mention the controls either, yet whose results are close to the final Table 5, where the controls are documented as included.

The concern is well-founded that baseline population could be a confounder. The cross-regency regression mentioned in section 2.1 shows that the number of new schools increased with subunitary elasticity with respect to population, meaning that more populous provinces received lower treatment per child. And since large regency population is correlated with being on the main island of Java—average regency population there was 1.08 million in 1995, against 0.41 million elsewhere—population is probably related to schooling and labor market outcomes for non-schooling reasons. In the 1970s and 1980s, Java became a locus of export-oriented industrialization, which affected individual schooling decisions, labor market experiences, and migration patterns (Heneveld 1979; Hill, Resosudarmo, and Vidyattama 2008).

However, it is not obvious that controlling away a known correlate of treatment improves estimation, since the treatment variation that remains is not necessarily more exogenous than that expunged. One can construct stories of confounding for any non-random component of treatment, such as the baseline primary school attendance rate, which is also controlled for in the intermediate and full control sets. Duflo (2004b) argues that in Duflo (2001), "identification is made possible because the allocation rule for schools is known." That point loses some force when one controls away the large, known components of allocation.

Ideally when exercising discretion in the face of such ambiguity, the decisions about which components of identifying variation to retain and which to control away are explained in text. And ideally they are decided before seeing the results. By linking the inclusion to the results, the working paper endnote reduces the certainty one can have that that was the sequence here. That makes it informative and minimally arbitrary to check the effect of dropping the controls.

### 2.5 Weak identification

Econometricians have made progress since 2001 in understanding and managing the effects of weak identification on IV estimators (Stock and Yogo 2002, Moreira 2003, Kleibergen and Paap 2006, Montiel Olea and Pfluger 2013). As identification weakens, IV biases toward OLS. Various tests have been developed for strength of identification. But screening on such pre-tests can itself generate bias and size distortions. Better, argues Andrews, Stock, and Sun (2018, §4.1), to favor estimators that are robust to weak identification. For forming confidence intervals, Finlay and Magnusson (2019)'s Monte Carlo–based review favors the Anderson-Rubin (AR, 1949) test with a wild-bootstrapped distribution for the test statistic (Roodman et al. 2019).[8] In comparison with the conditional likelihood ratio test of Moreira (2003), the AR test has the advantage of being defined for exactly- as well as over-identified regressions. The price of this generality is that the test compounds the hypothesis of primary interest—that coefficients on instrumented variables take given values—with a second—that the instruments are valid. Rejection on the test is therefore ambiguous. Of course, whichever null is the cause of a rejection, it speaks to causal interpretation of the results.

### 2.6 Endogenous sampling

The survey used for follow-up in Duflo (2001) employed a complex sampling design. It oversampled low-population regencies in order to produce comparably accurate statistics for all regencies.[9] Within regencies, it sampled urban and rural areas non-proportionally. Survey weights included in the data set allow for adjustment for this non-proportional sampling. The code Duflo sent me incorporates weights in the 2×2 DID causal estimates but not in the TWFE causal estimates. This leads to the question of how the decisions to weight one and not the other affects results. The original text does not discuss weights.

In general, there are two reasons to weight observations. One is to improve statistics as *descriptions*, i.e., to make them more representative for the original setting, in this case Indonesia of 1995. That rationale is not compelling here. Not incorporating the weights merely makes the results representative for a different, hypothetical population, one that may be no less relevant for testing theories or making decisions in other contexts. And if effects are fairly homogenous, as suggested above, weighted and unweighted results should not differ greatly. The other reason is to improve *causal estimates*. For example, if sampling is endogenous within a given estimation context, then incorporating weights can counteract selection on the basis of the outcome, a potential source of bias (Hausman and Wise 1981). However, incorporating survey weights typically increases the variance of estimators.

Appendix B reviews the conditions under which unweighted OLS consistently estimates the slope of the linear best fit in the population. Two sufficient conditions emerge. Let the variables $x$ and $y$ be the treatment and outcome and let the relationship of interest in the population be $y = \beta x + e$. And let $p$ be the probability that a given member of the population is sampled. Approximately speaking, the *purely exogenous* sufficient condition allows $p$ to be correlated with $x$ but not $e$, while the *purely endogenous* condition allows $p$ to be correlated with $e$ but not $x$. Sampling can be related to the observed or unobserved determinant of the outcome, but not both (Solon, Haider, and

[8] Simulations in Davidson and MacKinnon (2010) of the non-clustered but heteroskedasticity-robust analog also demonstrate the good size properties of the wild-bootstrapped AR test.

[9] Surbakti (1995, p. 19) states this motivation with respect to the contemporaneous socioeconomic survey (SUSENAS) rounds.

Wooldridge 2015). In fact, as spelled out in the appendix, an additional condition is required in each case. For example, in the purely endogenous case, some patterns of heteroskedasticity can still render unweighted OLS inconsistent.

Unlike in the theory just thumbnailed, the regressions of interest here include controls, notably the dummies for year of birth and regency of birth. To adapt the sufficiency conditions, we apply them to outcome and treatment after partialling out the controls. If subjects did not *migrate* within Indonesia, this step would eliminate nearly all the association of the sampling probability with *both* the treatment and the error. Why? There is no sign that within the Duflo (2001) samples, the sampling probability is related to age—little sign that, for example, that men in their 30s were oversampled relative to men in their 20s from the same regency.[10] And if the sampling probability is constant over time within regencies, then the variation in it would be a pure regional effect, which would effectively be absorbed by the regency fixed effects.

The catch in that argument lies in the distinction between regency of birth and regency of residence at follow-up. The Duflo (2001) regressions control for birth regency fixed effects. The sampling probability depends mainly on the place of residence at follow-up—urban or rural, in a low- or high-population regency. The two differ for migrants. Internal migration was an important feature of Indonesian development in this period (Hill, Resosudarmo, and Vidyattama 2008); indeed, the government promoted it to reduce overcrowding in Java, even as many people moved to cities for work. Hsiao (2024) finds that Inpres stimulated outmigration. If factors not controlled for affected both the outcomes and the decision to move, these could create a correlation between the sampling probability and the regression error terms. And if the sampling probability is related to both treatment and the error, unweighted OLS is inconsistent.

The literature takes two main approaches to testing whether a given unweighted OLS specification is consistent (Bollen et al. 2016). The *weight association* approach directly tests the sort of sufficiency conditions just mentioned—though to date, it has only been used to check the purely exogenous condition, not the purely endogenous one. One intuitive example is a test in Pfeffermann and Sverchkov (2007), regresses $p$ on $x$ and $y$.[11] Statistical significance of the coefficient on $y$ indicates that sampling is not purely exogenous. Likewise—though this is not contemplated in the original proposal—significance of the coefficient on $x$ indicates that sampling is not purely endogenous.

For an initial look at the effect of weighting on the Duflo (2001) results, I apply this technique to representative schooling and wage regressions, specifically OLS estimates with the minimal control set (Duflo 2001, row 1, column 1, and row 2, column 4, of Table 4, panel 1). I incorporate data corrections and cluster standard errors. The results suggest that in the schooling regression the weights are purely endogenous, with $p = 0.00$ for the coefficient on schooling and $p = 0.73$ for the treatment proxy $D_j$. Leaving aside the additional technical condition stated in appendix B, this result suggests that unweighted estimation of schooling impacts is consistent. But in the wage regression the weights appear neither purely exogenous nor purely endogenous ($p = 0.00, 0.04$), which suggests that unweighted estimation of wage effects is inconsistent.

[10] A weighted regression of age at follow-up on the sampling probability, restricted to the 1995 follow up and those aged 2–24 in 1974, and controlling for birthplace fixed effects, yields a $t$ statistic of –0.58.

[11] The test is proposed for a particular application—small-area estimation—but Wang, Wang, and Yan (2022) finds that it performs well in more generic Monte Carlo simulations.

The other main approach to assessing whether weighting is needed for consistency is *difference-in-coefficients*. It works in the tradition of Hausman (1982) testing, calculating the statistical significance of the gap between unweighted and weighted estimates (Pfeffermann 1993). Within the Hausman framework, unweighted OLS is the efficient estimator and weighted is the more-robust estimator.[12] Hausman testing avoids the error of checking only the purely exogenous sampling condition, when satisfaction of the purely endogenous condition could also avert inconsistency. I therefore opt for the Hausman approach below. Because errors are not assumed to be classical, I will use a bootstrap rather than analytical formulas to estimate the variance of the difference between the two estimates, which is the denominator of the test statistic (Cameron and Trivedi 2005, p. 378).[13] In particular, I will run clustered wild bootstraps without the null imposed, with 9,999 replications (Cameron, Gelbach, and Miller 2008; Roodman et al. 2019).[14]

When a Hausman test casts doubt on the null that the unweighted and unweighted estimates agree asymptotically, that can be explained in two ways. Treatment effects may be heterogeneous enough that weighting causes a large change in the estimated average effect. Cutting against that explanation would be the null result reported in appendix A from a check for effect heterogeneity. The alternative explanation is that the unweighted estimates are inconsistent.

One disadvantage of using inverse-sampling-probability weights is that their distribution can be skewed, making some observations inordinately influential. In the sample for the schooling regression just mentioned, the mean weight is 211, the 99$^{th}$ percentile is 772, and the maximum is 2,643. To limit the influence of extreme weights, I will winsorize them to the median weight plus 4 times the interquartile range. This is one simple approach evaluated through simulation in Potter and Zheng (2015).[15] In this case, it yields a limit of 903.

[12] A striking fact makes the connection to Hausman testing: weighted OLS is equivalent to unweighted 2SLS regression in which all the predictors $x_i$ are instrumented by their products with the inverse-probability weights, $wx_i$ (Winship and Radbill 1994, note 13). So comparing weighted and unweighted OLS can be reframed as comparing OLS and 2SLS.

[13] I use a nonparametric/pairs bootstrap. I draw whole clusters, with equal probability, and with replacement, with 1,000 replications. There has been little Monte Carlo testing of methods for assessing whether weighting is needed for consistency (Bollen et al. 2016). The exception is Wang, Wang, and Yan (2022), which validates the Hausman-type approach taken here as robust and competitively powered, but does not investigate bootstrap-based extensions to accommodate non-classical errors. In such contexts, the more-efficient (unweighted) estimator is not fully efficient, *contra* an assumption in the original definition of the Hausman test. Generalized least-squares, if feasible, would be efficient.

[14] Separate bootstraps are run for the unweighted and weighted regressions, The difference between the unweighted and weighted estimates is computed for each of the 9,999 bootstrap data sets. Then their variance is computed. The bootstraps for the two regressions are started with the same random number seed, to produce corresponding simulated data sets. For a modest boost in power, Davidson and MacKinnon (1999) favors imposing on the bootstrap data generating process the null that is being tested. That guidance is not applicable here because the null that the unweighted and weighted estimates agree cannot be imposed on either bootstrap alone.

## 2.7 Non-parallel trends

The geographic treatment variation in Inpres is not as good as random, for high- and low-treatment regencies presumptively differed at baseline. And it would be surprising if regencies that started out different would nevertheless have experienced the same *trends* in the absence of Inpres SD. In other words, the parallel trends assumption is almost surely violated (Kahn-Lang and Lang 2019). How much it is violated is unclear *a priori*. One common response to this sort of concern, exemplified by the Duflo (2001) placebo checks described in section 1, is to check for non-parallel trends in the pre-treatment era, and then rely on the experimental results if the pre-trend test result is not significant. But in simulations calibrated to a set of published DID studies, Roth (2022) concludes that half the time tests of this sort miss the presence of pre-trends that are large enough to cause substantial bias. The problem, as Kahn-Lang and Lang (2019) write, is that "not rejecting the null hypothesis is not equivalent to confirming it. Yet, in such settings, economists may be tempted to, in Brad DeLong's terms, 'seize the high ground of the null hypothesis,' which can lead to incorrect estimation of treatment effects." Roth (2022) also shows that filtering estimates for passage of a pre-trend pre-test itself induces bias, because of regression to the mean in the post-treatment period. As when facing weak identification, it appears less biased to estimate effects in a way that is robust to the problem than to filter results with a pre-test for it.

In the present context, we can concretely ground a concern about non-parallel trends. A common pattern is present in the Indonesia data. In their early 20s, educated wage workers earn a small multiple of what their less-educated peers earn. But the multiple rises with age. Figure 1 demonstrates. Each contour is a smoothed fit to the association between the log hourly wage and age, in a particular schooling stratum in 1995. All fits are restricted to wage-earning men and incorporate survey weights. As one scans from younger to older workers, the wage scale dilates with respect to schooling, even in logarithms. Mincer (1974) observes a similar pattern in U.S. data from 1960 and connects it to a theory of causation from schooling to earnings. The result is the Mincer labor function, which has the log wage linear in years of schooling and quadratic in years of work experience. The wage scale widens as one moves from the youngest to prime-age workers because more-educated people are later to enter the workforce, later to start accumulating experience, and later to hit diminishing returns to experience.

[15] This is a late change prompted by a comment from a referee. It increases consistency with the Roodman (2023) reanalysis of the Khanna (2023), in which the weight distribution is even more skewed.

The empirical fact of age-differentiated wage scale dilation provides the starting point for one story for non-parallel trends. Table 3, panel A, of Duflo (2001) documents that regencies that received higher treatment produced wage-earning men with less schooling on average: 8.02 versus 9.40 years in the pre-treatment cohorts (aged 12–17 in 1974) and 8.49 versus 9.76 in the post-treatment cohorts (aged 2–6). In the 1995 follow-up used in Duflo (2001), this negative selection causes natives of high-treatment areas to congregate disproportionately near the lower curves in Figure 1. Within this group, the before-after difference in wages—the wage drop as one moves from older to younger cohorts—should be relatively small, as the lower curves are flatter. Among workers from less-treated, more-schooled regencies, the wage drop should be larger, as their wage curves slope downward as one scans from right to left. The difference in those differences—the small, negative change for high-treatment regencies minus the larger, negative change for low-treatment ones—is positive. It should be expected to produce positive DID estimates even in the counterfactual of no effect from Inpres SD treatment.[16]

To explore potential biases from wage scale dilation in the 1995 follow-up, I plot the dilation of the wage scale at each age, defined as the slope from an age-specific OLS regression of the log hourly wage on schooling, among those with wage employment. The slopes and 95% confidence intervals are estimated in a single, omnibus regression, with weights and clustering. See Figure 2. From age 15 till about age 55, the wage scale widens with age. Within the Duflo (2001) follow-up, ages 23–45 as of 1995 (2–24 as of 1974), the pace of dilation is approximately linear, so some strategy to control for linear trends may remove most of the confounding.

Because Inpres treatment was roughly simultaneous in all regencies, if varying in intensity, it is relatively straightforward to model and control for pre-treatment trends.[17] The approach taken here formalizes an informal characterization in Duflo (2001) of the event studies. The paper's Figure 1 shows how the cross-sectional association between treatment and schooling attainment evolves, birth cohort by birth cohort. Figure 3 adds a similar plot for log wages. Of the first, the text observes that the "coefficients fluctuate around 0 until age 12 [as of 1974] and start increasing after age 12" (Duflo 2001, p. 801). A sustained trend break at age 12 in 1974 is plausible because Inpres began adding primary schools in 1974 and continued doing so through at least 1980, making the 11-year-olds of

[16] de Chaisemartin and D'Haultfœuille (2017, supplement 3.3.1), makes a similar point, starting from the Mincer model and equating treatment to schooling attainment. Here, the story has been extended by distinguishing Duflo (2001)'s treatment variable from schooling attainment and taking account of their negative cross-sectional association.

[17] Incorporating pre-trends is harder when treatment is staggered across cross-sectional units. See the debate over state-level no-fault divorce laws in Peters (1986), Wolfers (2006), and Lee and Solon (2011).

1974 approximately the first birth cohort young enough to attend the new schools. However, because of the uncertainty bands around the point estimates in Duflo (2001)'s Figure 1 (which will be widened by clustering), and because this observation of a change in trend is not formalized, it is not clear how easily the null of *no* change can be rejected.

To check for kinks, I will run regressions formulated by modifying $\widetilde{\mathbf{T}}_t$ in (3). $\widetilde{\mathbf{T}}_t$ will become a vector with two terms: a time trend $t$ (age in 1974) and a spline term that emerges for younger cohorts after age 12:

$$\max(0, 12 - t) \tag{4}$$

This allows a kink at age 12 in 1974. In OLS/reduced-form regressions, both components of $D_j\widetilde{\mathbf{T}}_t$ will enter as regressors. In 2SLS, the first will enter as a control while the second will instrument the schooling variable. I will restrict the model to ages 2–22, to extend each spline segment 10 years. This choice is minimally arbitrary and nearly matches the 2–23 range of Duflo (2001)'s Figure 1. Making the lengths multiples of five also minimizes distortions from age heaping, which is the tendency of poorer people to have their survey-recorded ages rounded to the nearest multiple of 5.

**Figure 1. Smoothed fits of hourly wage to age, by years of schooling, wage-earning Indonesian men in 1995**

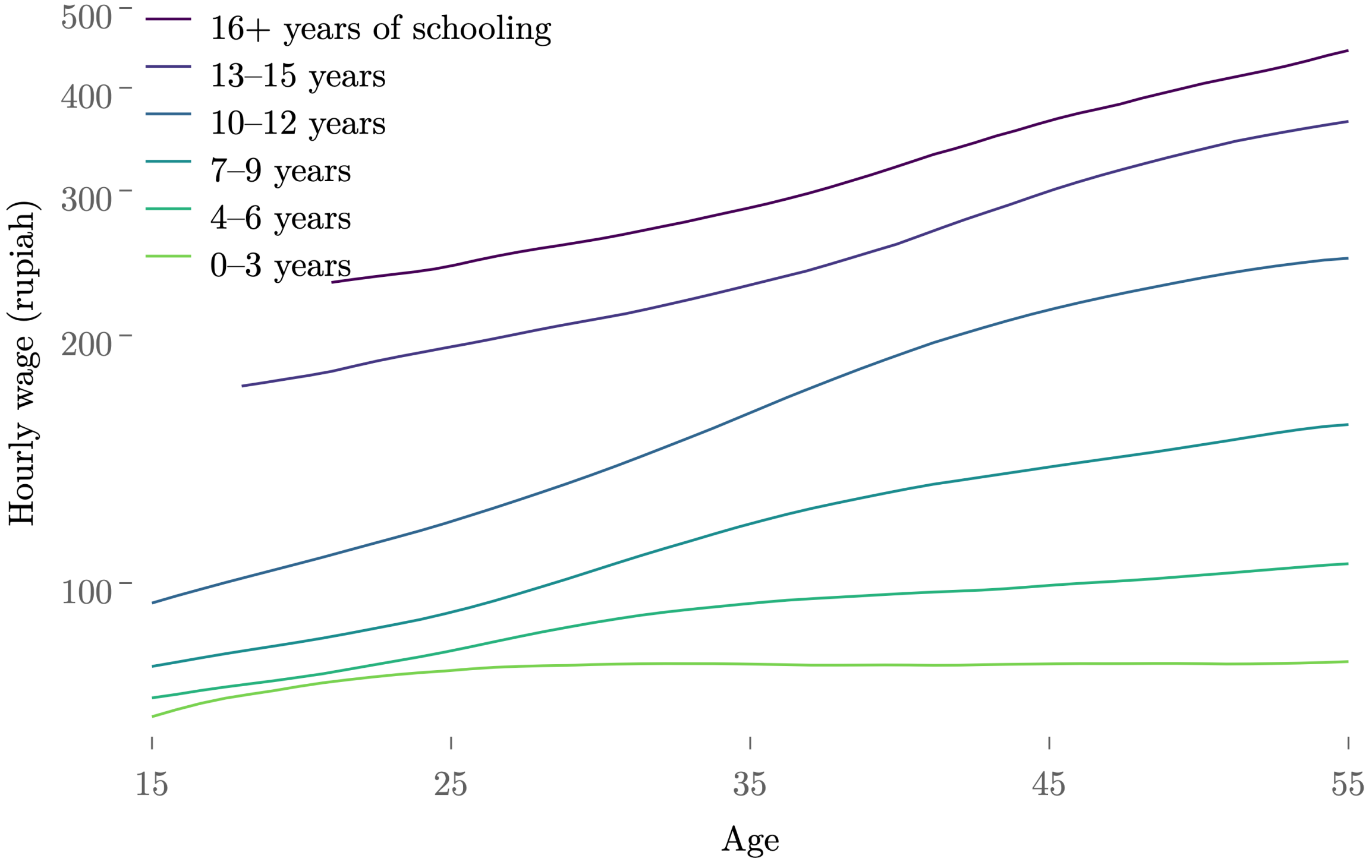

*Notes:* Curves are smooth local polynomial fits for the relationship between log wages and age, among men with wage work. Survey weights are incorporated.

**Figure 2. Wage scale dilation by age among Indonesian men in 1995**

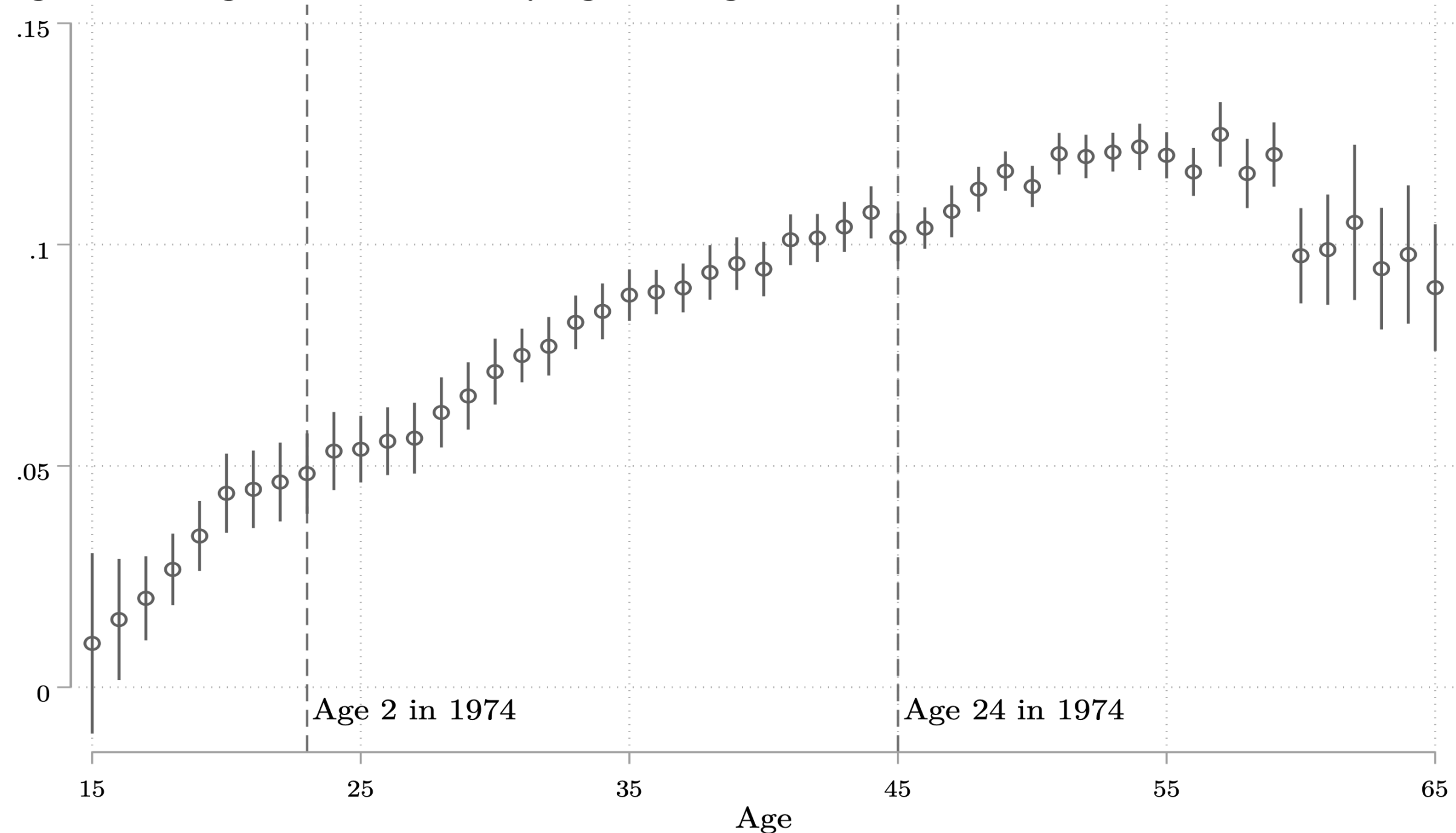

*Notes:* This graph shows the slope of the linear association between the log hourly wage and years of schooling for men at each year of age in the 1995 data. Survey weights are incorporated. The plotted 95% confidence intervals are clustered by birth regency.

# 3 Revising the Duflo (2001) causal estimates

## 3.1 2×2 DID

The reanalysis begins where the original analysis begins, with 2×2 DID estimates based on (1). A birth regency is "high treatment" if it garners a positive residual in a cross-regency regression of planned school construction on the number of children 5–14.

See Table 1 for the results. The first column contains the Duflo (2001) estimates. In the experiment of interest, which benchmarks outcomes for the 2–6-year-olds of 1974 against those for the 12–17-year-olds, the schooling impact is 0.12 years and the wage impact 0.026 log points. The ratio of those is the Wald DID estimate of the return to a year of schooling, 0.22; it is also shown in the column. In the placebo check, which compares the pre-treatment groups aged 18–24 and 12–17 as of 1974, the estimated "impacts" are hard to distinguish from zero: 0.034 for years of schooling and 0.0070 for the log hourly wage.

The *third* column of Table 1 shows the results from the data and code provided by Duflo, which differ from the published results.[18] The estimated impact on the log hourly wage falls from 0.26 to 0.16 and the return to schooling from 0.22 to 0.14. These estimates incorporate survey weights. For consistency with later tables, the *second* column shows the results when modifying the Duflo code to drop weights. This takes us further from the published results, which suggests that the published estimates indeed incorporate weights.

[18] Despite engagement with the original author, I have not managed to reconcile this discrepancy.

The last two columns document that the 2×2 results are not robust to correcting data errors. With or without weighting, impacts are now higher in the placebo check than in the experiment. The schooling impact is wrong-signed. That said, these contrasts are statistically weak, especially in light of the clustered standard errors now reported. The weighted Wald estimate of the return to schooling is –0.22, but is weakly identified and indistinguishable from zero. (Here, the weights are winsorized to the median weight plus 4 times the interquartile range, as discussed in section 2.6.)

**Table 1. 2×2 DID estimates of effects on schooling attainment on the log hourly wage, 1995**

| | Reported in Duflo (2001) | Duflo data & code | | New | |
|---|---|---|---|---|---|
| *Effect on schooling* | | | | | |
| Experiment | 0.12 | 0.10 | 0.11 | –0.080 | –0.029 |
| | (0.089) | (0.089) | (0.088) | (0.15) | (0.16) |
| Placebo | 0.034 | –0.037 | 0.0047 | 0.0030 | 0.072 |
| | (0.098) | (0.098) | (0.098) | (0.11) | (0.14) |
| *Effect on log hourly wage* | | | | | |
| Experiment | 0.026 | –0.0018 | 0.016 | –0.010 | 0.0065 |
| | (0.015) | (0.015) | (0.015) | (0.026) | (0.026) |
| Placebo | 0.0070 | 0.00083 | 0.0044 | 0.0047 | 0.014 |
| | (0.016) | (0.016) | (0.016) | (0.017) | (0.019) |
| *Return to schooling (Wald DID)* | | | | | |
| Experiment | 0.22 | –0.017 | 0.14 | 0.13 | –0.22 |
| | | (0.15) | (0.13) | (0.24) | (1.81) |
| Placebo | 0.21 | –0.023 | 0.93 | 1.55 | 0.19 |
| | | (0.48) | (17.5) | (55.2) | (0.30) |
| Data corrections | | | | ✓ | ✓ |
| Clustered | | | | ✓ | ✓ |
| Weighted | ✓? | | ✓ | | ✓ |

Notes: Sample is restricted to wage earners aged 12–17 and 18–24 in the placebo experiment and 2–6 and 12–17 in the experiment of interest, with the younger range in each case being the post-treatment group. Wald DID estimates are ratios of corresponding results in first two panels. The corrected high-treatment indicator is constructed by returning to primary sources for planned school construction and student populations, then defining high treatment as having a positive residual in a linear regression of the former on the latter (Duflo 2001, note 2). "Clustered" means clustered by regency of birth. The Duflo code does not run unweighted 2×2 DID regressions, but is here modified to do so. In the last regression, weights are winsorized to the median weight plus 4 times the interquartile range.

## 3.2 OLS/reduced-form results

### Years of schooling

I move next to OLS TWFE estimates, which are much more central to Duflo (2001). Samples are restricted to the same cohorts as in the 2×2 DID. Fixed-effect dummies are included for birth year and birth regency. In the original, weights are now dropped. Here, the original estimates are exactly reproduced.

I focus first on the educational outcome, years of schooling. Table 2 displays the original results along with the consequences of introducing changes to address issues raised in section 2. The results are arranged so that the original appears in the upper left and the one incorporating all changes is in the bottom right.

In panel A, the minimal Duflo (2001) control set is included; recall that it consists of interactions between cohort dummies and the number of children aged 5–14 in the regency of birth. In the original, adding controls beyond the minimal set hardly affects results, so for parsimony those specifications are omitted. As shown in the upper left of the panel, Duflo (2001) estimates that Inpres increased schooling attainment by 0.12 years (standard error 0.025) for each planned new school per 1,000 children. The corresponding placebo estimate is also reported. However, unlike in Duflo (2001), it is rescaled for comparability with the experimental effect. The reasoning runs as follows. The average lag between the two placebo group age ranges—between ages 18–24 and 12–17 as of 1974—is the distance between their midpoints, $21 - 14.5 = 6.5$ years. The placebo estimates are therefore of cumulative trends over 6.5 years. The experimental comparison—between the 12–17 and 2–6 groups—spans an average gap of $14.5 - 4 = 10.5$ years. Comparability calls for rescaling the placebo results by 10.5/6.5. Despite this magnification, in the original specification, the placebo effect is estimated to be just 0.015 years per unit of treatment, barely a tenth of the experimental effect (panel A, column 1).

Moving to column 2, we see that unlike in 2×2 DID, correcting the data errors barely perturbs the point estimates. However, clustering the standard errors increases them by about 50%.

Weighting is introduced in column 3 of panel A. It too increases standard errors, though more modestly. It also makes the placebo impacts bigger than the impacts of the experiment of interest: the main estimate drops from 0.12 (standard error 0.042) to 0.077 (0.054) while the rescaled placebo effect rises from 0.013 (0.055) to 0.083 (0.068). The $p$ values from the Hausman tests comparing weighted and unweighted results, reported in the last column of panel A, cast these shifts as unlikely if treatment effects are homogeneous and unweighted OLS is consistent. The $p$ value is 0.23 for the experimental effect and 0.080 for the placebo effect. And for specification tests, a low significance standard such as $p < 0.05$ is the opposite of conservative. The weighted findings challenge the assumption that pre-treatment schooling trends are close enough to parallel that, if continued into the post-treatment period, they could explain only a small fraction of the measured effect.

The bottom row of panel A reports the results from the preferred method for controlling for regency-specific linear trends, using the kink model. Recall that this fits a spline with two 10-year segments connected at age 12 in 1974. These estimates are also depicted in the plots of Figure 3, which are event studies like Duflo (2001, Figure 1) with the kink fits superimposed. For example, in the specification closest to that in Duflo (2001), but subject to data corrections and clustering, the trend in years of schooling jumps steepens by an estimated 0.0072 per year at the allowed kink point; see the upper left plot of the figure. Following that steeper trend 8 years to the right from the allowed kink at age 12 brings us to the center of the age 2–6 range, which is the post-treatment cohort. The kink implies that the 2–6 cohort averaged $8 \times 0.0072 = 0.058$ more years in school for each unit of treatment—more, that is, than would have been the case with no deviation from the pre-trend. This kink-based estimate is displayed in the figure and in panel A along with its clustered standard error, 0.071. The corresponding weighted estimate appears to the right in both displays; it is statistically similar.[19]

One specification change has been held back to this point: dropping the Duflo (2001) controls based on child population. Section 2.3 laid out reasons to be somewhat skeptical of their inclusion. In addition, when estimating impacts from distinctive kinks in long-term trends, as preferred here, extra controls are in principle not needed for consistency, even as they consume statistical power. Aside from its connection through Inpres SD, we have little reason to expect initial population to create its own kink that would mask the true effect unless controlled for—little reason, that is, to expect the time-varying impact of initial population to break distinctively at the cohort aged 12 in 1974.

For these reasons, the bottom halves of Table 2 and Figure 3 show the results from rerunning the preceding analysis without the extra controls. (The results in the "original" column of panel B do not appear in Duflo (2001) and are obtained by modifying the corresponding regressions in panel A.) Precision slightly increases. The experimental effect barely falls and the placebo effect rises further. Going by the Hausman $p$ values, weighted and unweighted estimates now better agree. The kink in the long-term schooling trend is even closer to zero, indeed slightly negative.

If instead of dropping these controls, they are retained in logarithms, the results are more favorable to the original. The schooling impact estimates are somewhat larger, the placebo effects somewhat smaller, and the kink somewhat sharper. (See appendix C.) Still, the Hausman test grounds skepticism of the unweighted results, and in the weighted results the placebo effect is almost as large as the experimental effect. And the evidence of a schooling kink remains ambiguous; it is estimated at 0.094 (0.074) without weights and 0.072 (0.090) with.

[19] Switching from the minimal to the intermediate control set as in Duflo (2001, Figure 1)—adding interactions between the baseline enrollment rate and birth year dummies—yields a kink estimate of 0.054 (0.075) with clustering and 0.054 (0.063) without. The latter result corresponds most directly to the original figure, but incorporates data corrections. This test provides little evidence of a trend break.

In sum, it is hard to be confident on the basis of the 1995 data that Inpres SD caused a break from long-term schooling trends. As one reverses choices that are not explained in the original text, and that are applied to these specifications and not some others, the treatment effect falls, the placebo effect rises, and the suggestion of a trend break fades.

**Table 2. Reduced-form TWFE impact estimates for schooling, 1995**

| | Original | New | | |
|---|---|---|---|---|
| | (1) | (2) | (3) | Hausman *p*: weighted vs. not |
| *A. With controls based on number of children aged 5–14* | | | | |
| Experiment | 0.12 | 0.12 | 0.077 | 0.23 |
| | (0.025) | (0.042) | (0.054) | |
| Placebo | 0.015 | 0.013 | 0.083 | 0.080 |
| | (0.042) | (0.055) | (0.068) | |
| Kink | | 0.058 | 0.040 | 0.73 |
| | | (0.071) | (0.087) | |
| *B. Without extra controls* | | | | |
| Experiment | 0.10 | 0.100 | 0.075 | 0.45 |
| | (0.023) | (0.041) | (0.050) | |
| Placebo | 0.058 | 0.063 | 0.11 | 0.25 |
| | (0.038) | (0.052) | (0.068) | |
| Kink | | –0.016 | –0.027 | 0.83 |
| | | (0.064) | (0.082) | |
| Data corrections | | ✓ | ✓ | |
| Clustered | | ✓ | ✓ | |
| Weighted | | | ✓ | |
| Experiment *N* | 78,470 | 78,470 | 78,470 | |
| Placebo *N* | 78,488 | 78,488 | 78,488 | |
| Kink *N* | | 143,143 | 143,143 | |

*Notes:* Coefficient estimates are for $D_j T_t$ where $D_j$ is planned Inpres schools per 1,000 children and $T_t$ is a dummy for the younger cohorts in each experiment. Hausman *p* is for the Hausman test of the unweighted estimate against the weighted, using a 9,999-replication wild cluster bootstrapped estimate of the covariance matrix for the difference between the two. In the last regression, weights are winsorized to the median weight plus 4 times the interquartile range. Experiment samples are restricted to men aged 2–6 or 12–17 in 1974 and placebo samples to ages 12–17 and 18–24. "Original" results without extra controls are obtained by modifying regressions that exactly replicate the original. All regressions control for survey year, year-of-birth, and regency-of-birth dummies. Standard errors are in parentheses—classical in the original estimates and clustered by birth regency in the new. Since the average age difference between treatment and control is 10.5 years in the experiment and 6.5 in the placebo test, for comparability, the placebo results are scaled by 10.5/6.5. Kink estimates are coefficients on the kink term $D_j \times \max(0,12-t)$, multiplied by 8, which is the average lag from the allowed moment of kink, age 12 in 1974, to the treatment group aged 2–6 in 1974.

**Figure 3. Coefficients on interactions between age dummies and Inpres intensity in schooling regressions, 1995**

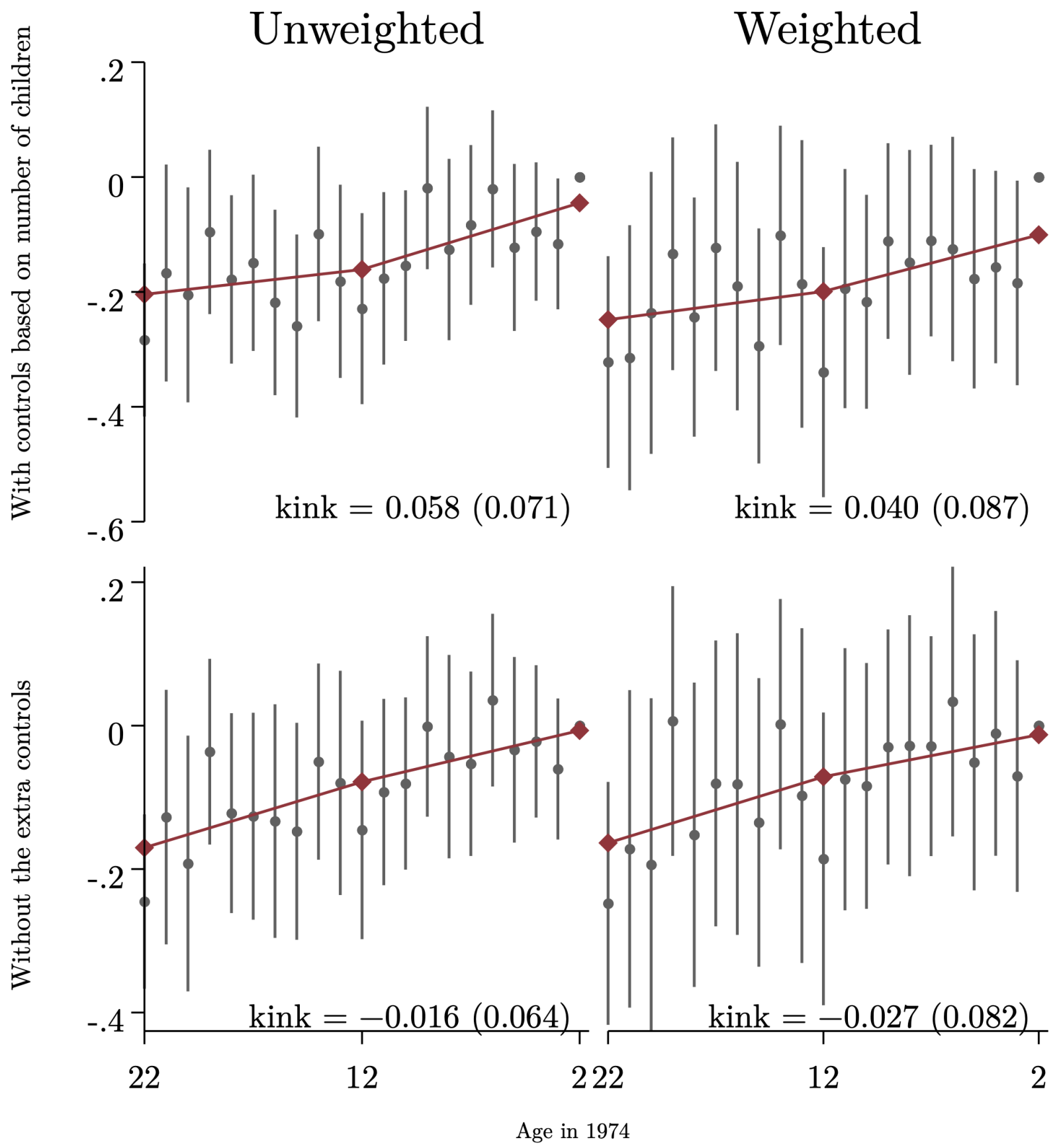

*Notes:* Each plot depicts, in grey, the coefficient estimates and 95% confidence intervals for interactions between birth year cohort dummies and the treatment intensity indicator, in regressions reported in Table 2, panel A. Age 24 in 1974 is the omitted reference cohort. Controls based on child population are products of the elementary school–age population in regency of birth and age cohort dummies. Kink model fits are superimposed.

**Reduced-form labor impacts**

We turn next to reduced-form impacts on the labor outcomes—namely, employment (wage sector participation) and the logarithm of the hourly wage. Although the original does not report reduced-form employment estimates, they are covered here. The analysis proceeds exactly as before: see Table 3 and Figure 4. With one striking exception, the story is similar. The original reduced-form wage results are exactly reproduced. Clustering and weighting reduce precision. Dropping the extra controls increases precision modestly, even as it reduces the kink estimates.

For employment, most impact estimates are positive and reasonably significant. The estimate from the specification closest to the original, 2.1 (0.44) points of employment per unit of treatment, becomes 1.7 (0.63) when weighting and 1.2 (0.60) when also dropping the extra controls. The kink estimates are similar without weights, but shrink with weighting. The Hausman test tends to view the differences between unweighted and weighted estimates as more than chance. (See columns 1 and 2 of Table 3 and Figure 4).

The "striking exception" to the tendency for the revisions to weaken results is a substantial strengthening of the apparent wage impact when relying on the kink model. On the one hand, dropping the child population–based controls from the wage specification substantially alters the original impact estimates, from 0.015 (0.0072) log points per unit of treatment to –0.014 (0.0063) (Table 3, column 3). This confirms the Duflo (2000, note 21) statement that the controls are important for the wage equation. On the other hand, the pre-trend is negative in all four variants, as seen in the right half of Figure 4. Against that benchmark, even a relatively flat post-treatment trend in the event study constitutes a large, positive trend break. And the kink is about twice as large with weighting, so that the Hausman tests leans against the null that the weighted and unweighted estimators are asymptotically equivalent. With weights and without the extra controls, the kink-based impact estimate is 0.057 (0.024) log points per unit of treatment. That is almost four times the reduced-form wage impact of 0.015 reported in Duflo (2001).

The finding of a strong trend break attributable to Inpres, in wages but not schooling, poses a puzzle. It should somewhat increase our credence in the proposition that Inpres *did* increase wages via schooling, in a way that the schooling regressions are failing to detect, perhaps because of noisy measurement. While the kink-based results in Table 2 do not allow one to reject the null of no impact on schooling, neither do they make it easy to reject, say, the Duflo (2001) estimate of 0.12 years of extra schooling per new school per 1,000 children.

**Table 3. Reduced-form TWFE impact estimates for labor outcomes, 1995**

| | Formal sector employment | | | Log hourly wage | | | |
|---|---|---|---|---|---|---|---|
| | New | | | Original | New | | |
| | (1) | (2) | Hausman $p$: weighted vs. not | (3) | (4) | (5) | Hausman $p$: weighted vs. not |
| *A. With controls based on number of children aged 5–14* | | | | | | | |
| Experiment | 0.021 | 0.017 | 0.35 | 0.015 | 0.019 | 0.018 | 0.95 |
| | (0.0044) | (0.0063) | | (0.0072) | (0.011) | (0.016) | |
| Placebo | 0.0011 | 0.0079 | 0.19 | 0.0051 | –0.0046 | –0.0065 | 0.88 |
| | (0.0060) | (0.0078) | | (0.013) | (0.013) | (0.018) | |
| Kink | 0.022 | 0.011 | 0.14 | | 0.035 | 0.065 | 0.060 |
| | (0.0079) | (0.011) | | | (0.018) | (0.026) | |
| *B. Without extra controls* | | | | | | | |
| Experiment | 0.013 | 0.012 | 0.75 | –0.014 | –0.014 | –0.011 | 0.69 |
| | (0.0044) | (0.0060) | | (0.0063) | (0.0096) | (0.014) | |
| Placebo | 0.0014 | 0.0053 | 0.42 | –0.0070 | –0.017 | –0.023 | 0.63 |
| | (0.0057) | (0.0074) | | (0.011) | (0.012) | (0.017) | |
| Kink | 0.015 | 0.0059 | 0.15 | | 0.032 | 0.057 | 0.075 |
| | (0.0082) | (0.010) | | | (0.016) | (0.024) | |
| Data corrections | ✓ | ✓ | | | ✓ | ✓ | |
| Clustered | ✓ | ✓ | | | ✓ | ✓ | |
| Weighted | | ✓ | | | | ✓ | |
| Experiment $N$ | 78,470 | 78,470 | | 31,061 | 31,061 | 31,061 | |
| Placebo $N$ | 78,488 | 78,488 | | 30,458 | 30,458 | 30,458 | |
| Kink $N$ | 143,143 | 143,143 | | | 57,234 | 57,234 | |

*Notes:* See notes to Table 2.

**Figure 4. Coefficients on interactions between age dummies and Inpres intensity in labor outcome regressions, 1995**

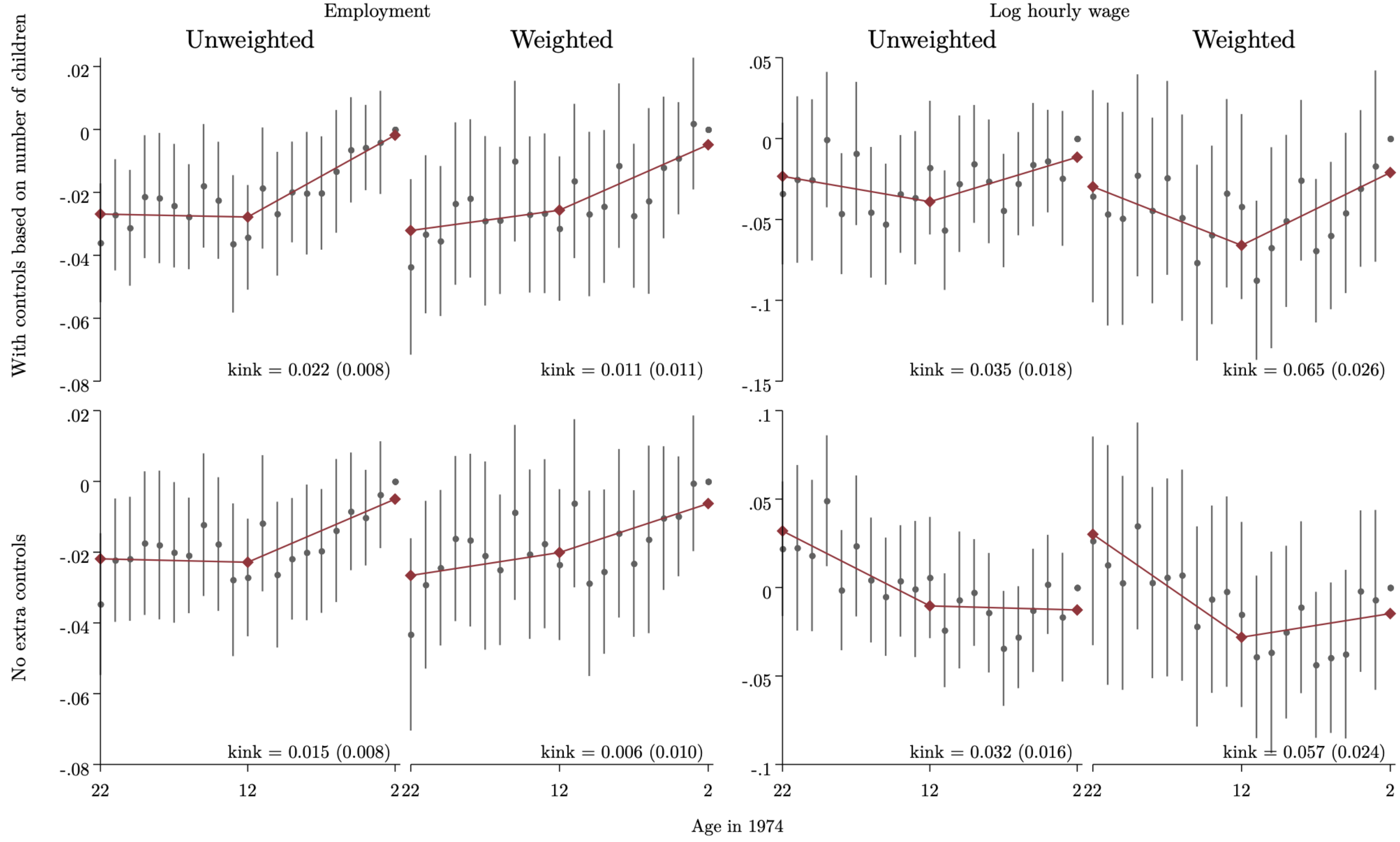

*Notes:* See notes for Figure 3.

### 3.3 2SLS estimates of the returns to schooling

I finish the reanalysis of the 1995 follow-up with 2SLS estimates of the returns to schooling. As noted, the preceding results are compatible with the hypothesis that Inpres SD boosted schooling and thereby wages. However, the weakness of the schooling results suggest that it will be difficult to measure the second causal step, the return to schooling.

As explained in sections 1 and 2.7, three instrument sets of the form $D_j\widetilde{\mathbf{T}}_t$ are tried, where $D_j$ is treatment intensity in regency $j$. The first two sets are from Duflo (2001): the exactly identified "by young/old" specification, in which $\widetilde{\mathbf{T}}_t$ consists solely of a dummy for the group aged 2–6 in 1974; and the overidentified "by birth year" specification, with one dummy for those aged 12–24 and separate dummies for each yearly age in 2–11. The second appears preferred in Duflo (2001) because results from the first are reported only once. The third variant is new. It incorporates the kink model. $D_jt$ is added as a control and $D_j \times \max(0,12-t)$ becomes the instrument capturing slope change at the allowed kink point.

Once again, the original results are exactly reproduced. And once again, I correct baseline variables, cluster standard errors by birth regency, and, in various combinations, weight by inverse sampling probability and drop the controls based on child population. I also report the Kleibergen-Paap (KP, 2013) measure of identification strength, which was not available when Duflo (2001) was written.[20] Results for the "by young/old" and kink instrument set appear in Table 4. The estimates with by-birth-year instruments are in appendix Table D–1.

Instrument weakness proves prevalent. Only the specifications closest to the original (unweighted, with child population controls, using the original's less-preferred "by young/old" instrument), earn KP $F$ statistics above 7 (Table 4, panel A, columns 1 and 6), where 10 is a rule-of-thumb minimum (Andrews, Stock, and Sun 2018). The KP statistics for the by-birth-year instrument set never clear 2, which is why, for parsimony, they are moved to appendix D.[21] For the kink specifications (columns of 3, 4, 8, and 9 of Table 4), identification is similarly weak, presumably because measuring a trend change is more demanding than measuring a trend.

As mentioned in section 2.5, recent scholarship argues that filtering IV estimates on the basis of a pre-test introduces bias. Instead of putting all inferential weight on the results with, say, KP $F >$ 5, I engage in weak identification–robust inference for all of the estimates. To this end, Table 4 reports 95% confidence sets constructed using the wild-bootstrapped Anderson-Rubin test. Most of these confidence sets are disjoint or unbounded. Still, a confidence interval of $(-\infty, \infty)$ does not mean that the regression extracts no information. To provide insight, Figure 5 displays the associated confidence curves. These are functions of trial values for the return to schooling. For a given trial value, such as a return to schooling of 10%, the test statistic's distribution is simulated using the Wild Restricted Efficient bootstrap of Davidson and MacKinnon (2010).[22] The two-sided $p$ value extracted from the bootstrap is then plotted against the trial coefficient. The curve's intersections with $p = 0.05$ are pinpointed in order to demarcate the confidence sets. The plots are arrayed to parallel the presentation in Table 4.

[20] The test of Montiel Olea and Pflueger (2013) is in principle superior here because it is robust to heteroskedasticity and within-cluster dependence. However, it coincides with the KP $F$ in the exactly identified regressions that emerge as preferred here, produces comparably low values for the overidentified regressions, and is less practical because the available software is less well developed.

[21] It is counterintuitive that adding instruments weakens their collective strength. The cause of the paradox is that identification strength is, broadly, the ratio of two competing factors: the ability of the instruments to explain the instrumented variables, and their expected ability to explain the first-stage error, i.e., the endogenous component of the explanatory variables. The first is desirable; the second is not, since it generates bias toward OLS. But, mechanically, the second also rises with the instrument count. Evidently, moving to the larger instrument set is increasing the undesirable factor more.

Though the 95% confidence intervals are often unbounded, the original Duflo (2001) estimates sometimes remain recognizable within them. When the controls based on child population are included, the test assigns lower $p$ values to negative rates of return in both employment and wages (upper half of the figure).

These final results from the reanalysis of the 1995 data studied in Duflo (2001) once more produce ambiguity. They weaken the confidence we can have that more schooling led to higher formal-sector pay. The cluster-robust 95% confidence interval for the return to schooling in the specification closest to the original, at the top of column 6 in of Table 4, is $[-0.03, 0.32]$. While that rejects the null of no return to schooling at a low $p$ value, it does so on the basis of some identifying assumptions that are plausibly violated, including that eschewing weights does not cause bias, and that ambient trends are parallel. Both assumptions are challenged by earlier results: the Hausman tests for the first stage (Table 2, final column) lean against the equivalency of the unweighted and weighted estimators, while the kink-based wage analysis has shown that pre-trends are in general not negligible.

In the hope of bringing more precision, we move next to larger follow-ups in the prime working years.

[22] 99,999 replications are performed for each test. Auxiliary weights are Rademacher-distributed and drawn by regency. See Roodman et al. (2019).

**Table 4. 2SLS estimates of the effect of years of schooling on labor outcomes, 1995**

| | Employment | | | | Log hourly wage | | | | |
|---|---|---|---|---|---|---|---|---|---|
| | Instrument by young/old | | Kink instrument | | Instrument by young/old | | | Kink instrument | |
| | (1) | (2) | (3) | (4) | (5) | (6) | (7) | (8) | (9) |
| *A. With controls based on number of children aged 5–14* | | | | | | | | | |
| Coefficient | 0.17 | 0.22 | 0.35 | 0.36 | 0.0752 | 0.12 | 0.12 | 0.17 | 0.19 |
| | (0.065) | (0.16) | (0.38) | (0.71) | (0.0338) | (0.058) | (0.094) | (0.11) | (0.098) |
| | [0.09, 0.58] | (−∞, −0.43] ∪ [0.05, ∞) | (−∞, −0.28] ∪ [0.07, ∞) | (−∞, ∞) | | [−0.03, 0.32] | [−0.38, 1.04] | (−∞, ∞) | [0.00, ∞) |
| KP $F$ | 7.89 | 2.08 | 0.92 | 0.28 | | 7.44 | 4.29 | 2.22 | 4.23 |
| *B. Without extra controls* | | | | | | | | | |
| Coefficient | 0.13 | 0.16 | –1.73 | –0.38 | | –0.50 | –0.41 | 1.48 | 0.43 |
| | (0.063) | (0.11) | (9.97) | (0.93) | | (1.15) | (1.38) | (13.3) | (0.54) |
| | [0.04, 0.69] | (−∞, −0.24] ∪ [−0.00, ∞) | (−∞, −0.07] ∪ [0.08, ∞) | (−∞, ∞) | | (−∞, 0.07] ∪ [0.20, ∞) | (−∞, ∞) | (−∞, ∞) | (−∞, ∞) |
| KP $F$ | 5.96 | 2.21 | 0.030 | 0.21 | | 0.28 | 0.16 | 0.011 | 0.45 |
| Data corrections | ✓ | ✓ | ✓ | ✓ | | ✓ | ✓ | ✓ | ✓ |
| Clustered | ✓ | ✓ | ✓ | ✓ | | ✓ | ✓ | ✓ | ✓ |
| Weighted | | ✓ | | ✓ | | | ✓ | | ✓ |
| Pre-trend control | | | ✓ | ✓ | | | | ✓ | ✓ |
| Observations | 78,470 | 78,470 | 152,989 | 152,989 | 31,061 | 31,061 | 31,061 | 60,663 | 60,663 |

Notes: Each column of each panel shows results for a different dependent variable. Coefficients are for years of schooling. “Young/old” regressions are restricted to ages 2–6, 12–17. The kink specification adds $D_j t$ as a control and instruments with $D_j \times \max(0,12-t)$ where $t$ is age in 1974. All regressions control for year-of-birth, and regency-of-birth dummies. Standard errors are in parentheses and 95% confidence sets are in brackets. The latter are from the Anderson-Rubin test with a wild-bootstrapped distribution with 99,999 replications. “KP $F$” is the Kleibergen-Paap measure of instrument strength. Column 5 is copied from Duflo (2001), Table 7.

**Figure 5. Confidence curves from wild-bootstrapped tests of the coefficient on years of schooling in 2SLS regressions in Table 4**

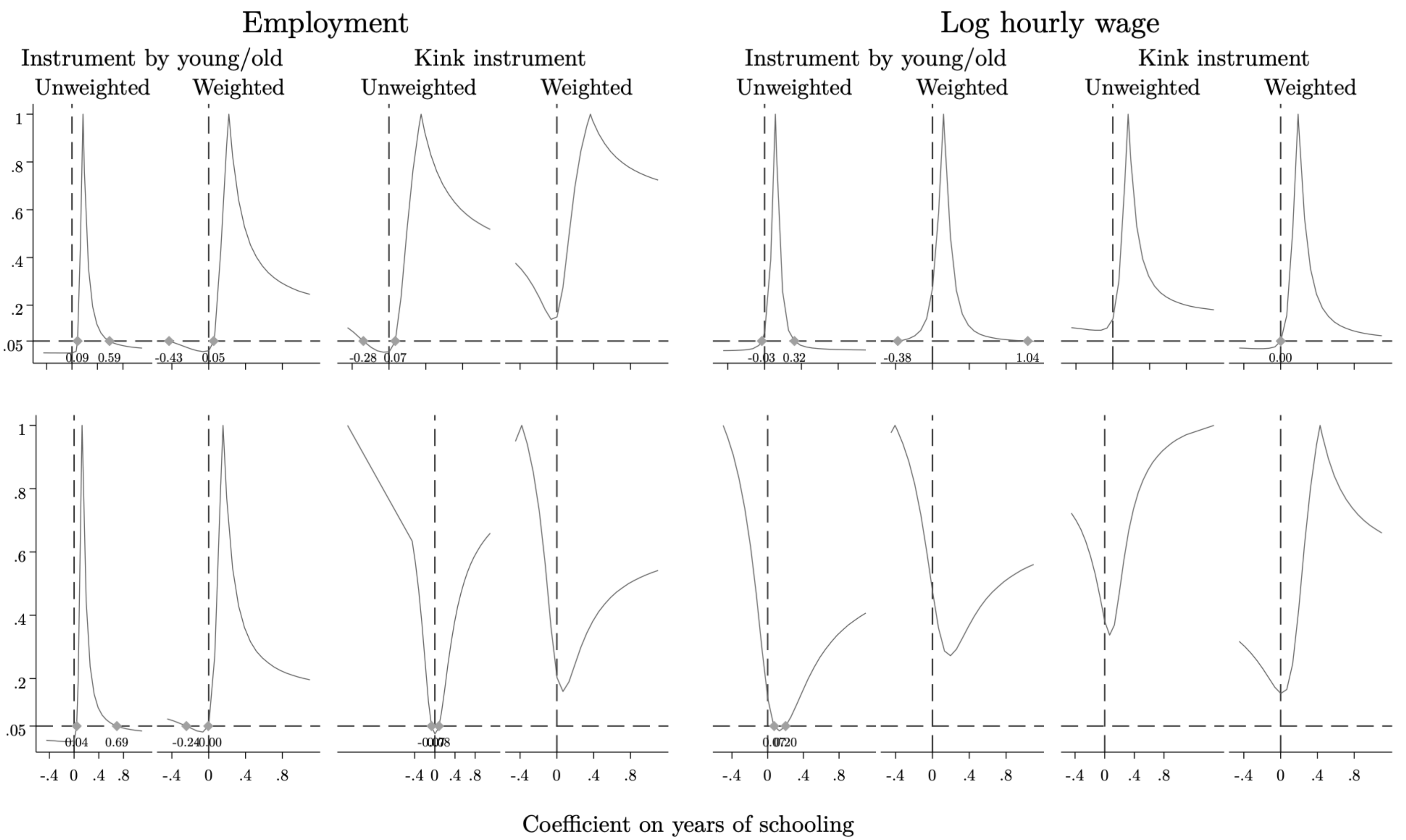

*Notes:* Each plot is of the $p$ value for the null that the instruments in a particular regression are valid and the coefficient on schooling takes a given trial value, as a function of that trial value. Each test statistic is from an Anderson-Rubin test, whose distribution is simulated using a wild bootstrap. Horizontal dashed lines mark $p = 0.05$. Their intersections with the curves define 95% confidence intervals. The plots are arranged in parallel with the results in Table 4, to which they correspond.

# 4 Following up later

## 4.1 The added surveys

The Indonesian statistical agency has long conducted several national surveys, with themes including labor and household welfare. As a result, the passage of time has generated more data on the subjects of the quasi-experiment framed in Duflo (2001). Since schooling attainment hardly changes during adulthood, the new surveys should help us to better measure the schooling impacts already studied in the 1995 data. In addition, information on labor market outcomes in prime work years is of interest in its own right. In the 1995 follow-up, the main treated cohorts are 2–6 years old in 1974, thus 23–27 at survey, early in their careers.

The choice of follow-ups is constrained by the need to construct key variables in Duflo (2001): regency and year of birth, years of schooling, employment, and wages. The rarest of those is regency of birth. Feasible follow-ups include:

- *The 2005 intercensal survey (SUPAS).* The successor to the 1995 survey, this one provides all the needed variables except wages.[23] It is hosted by the IPUMS project (MPC 2020).
- *Socioeconomic survey (SUSENAS).* The rounds of 2011–14 and 2017–19 cover schooling. The 2011–14 SUSENAS waves also include standard labor survey questions.[24]

One complication arises in moving to these later surveys: the possibility of selective attrition. This could result from mortality if, for example, subjects from more-treated areas are poorer and die sooner. In addition, historically, the standard retirement age for civil servants in Indonesia was 56. This may have caused those with formal-sector employment to exit the work force earlier. (A common formal profession in the data set is schoolteacher.) Duflo (2001)'s pre-treatment cohort, men aged 12–17 in 1974, started to turn 56 in 2013.

The attrition biases might not be large. In Indonesia in this period only tertiary schooling—not the target of Inpres—is strongly associated with mortality (Sudharsanan et al. 2020; Xu et al. 2024). And the progression of the pre-treatment cohorts into retirement may have been more complex and gradual than the existence of an official retirement age would imply. Government employees could work beyond age 56 under some circumstances (Muliati and Wiener 2014). And civil servants with at least 20 years of service could retire as early as 50. To the extent that attrition from retirement is gradual, it becomes just one more ambient trend that the statistical models are more or less designed to adjust for, e.g., with the linear time trend in the kink model.

[23] Previous versions of this paper imputed wages in 2005 with a model calibrated to the 1995 SUPAS data. This model was inspired by the income scores IPUMS has constructed for U.S. census data to proxy for earnings. But the validity of estimates based on imputed wages could be questioned. In fact, *all* wage indicators contain measurement problems (see Surbakti 1995, p. 44, on issues in wage values in Indonesia's surveys). Still, I dropped the imputation because I had developed it before I learned about the option of using later SUSENAS data. I probably would never have imputed the 2005 wages if I had known about the later data. Skipping the imputation reduces my exercise of researcher discretion. For much the same reasons, I also dropped the use of 2010 labor survey (SAKERNAS) data, despite the precedent in Duflo (2004a)'s use of the same survey to estimate medium-run impacts of Inpres. The data have the disadvantage of including regency of residence but not regency of birth.

[24] Hsiao (2024) first used the 2011–14 rounds to revisit Duflo (2001).

This matter can be probed empirically. If the bias is severe, then as pre-treatment cohorts (the 12–24-year-olds of 1974) progress through the surveys of 2011–19, the association *within* this demographic between treatment intensity in place of birth and outcomes later in life will shift as they age. For example, if more-treated regencies were poorer before Inpres and produced fewer well-paid, formal sector workers, then retirement of such workers could flatten the cross-sectional dependence of wages on treatment intensity. To check, I regress the outcomes of interest on interactions between treatment intensity and survey year dummies. Much as in the main TWFE specifications studied here, birth year, birth regency, and survey year fixed effects are controlled for. Observation weights are again incorporated, while controls based on child population are not.[25] Samples are restricted to the pre-treatment 12–24-year-olds. The coefficient estimates are plotted in Figure 6 along with clustered 95% confidence intervals. The last available year for a given outcome is always the omitted reference year, so it is represented with an exact zero. To the extent that the tests are powered to detect selective attrition, they do not find it. Except for employment, the confidence intervals easily embrace zero. And the positive relative trend in employment starts in 1995–2005 and thus is not strongly linked to the end of (work) life.[26]

Since attrition bias does not appear to be large, especially in the key schooling and wage variables, I will retain all the feasible follow-ups in the 2010s.

[25] Reversing these two choices make no difference for the conclusion.

[26] A previous version of this paper stopped follow-up at 2012 because the 12–17-year-olds of 1974, the Duflo (2001) pre-treatment cohorts, began reaching age 56 in 2013. Through the analysis described here, I changed my mind.

**Figure 6. Coefficients on interactions between survey year dummies and Inpres intensity, among 12–17-year-olds of 1974**

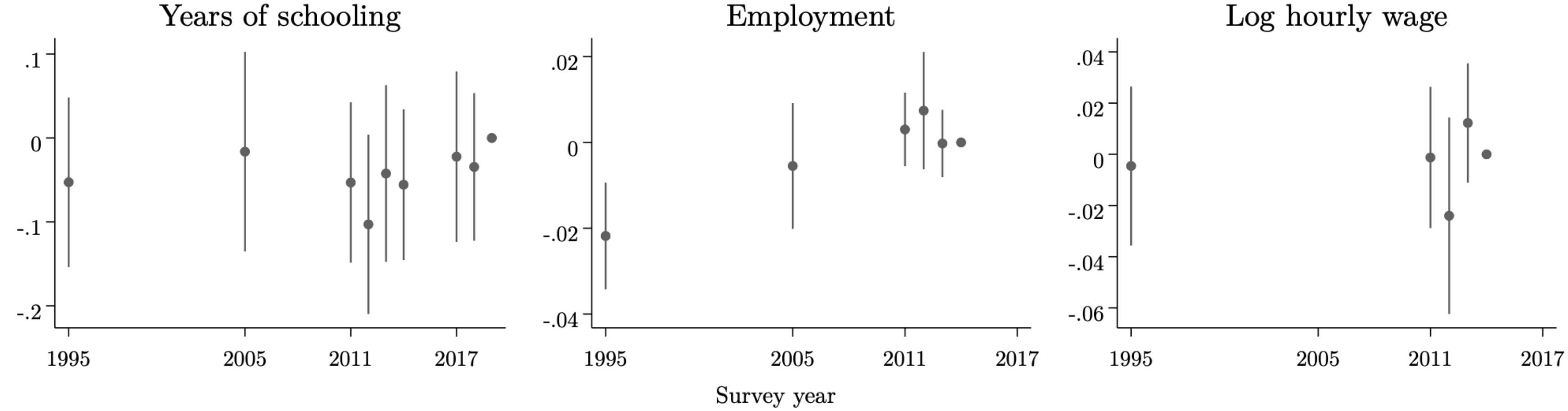

*Notes:* Each plot shows coefficients from a single regression, on the products of treatment intensity in regency of birth with survey year dummies. Each dependent variable is available in a distinct subset of survey rounds. Observation weights are incorporated. Samples are restricted to the pre-treatment cohorts aged 12–24 in 1974. All regressions control for birth year, birth regency, and survey year fixed effects. The last available year is always the omitted reference year, so it is represented with an exact 0.

## 4.2 Results from later follow-ups

To provide an out-of-sample test, the TWFE specifications are transferred from the 1995 follow-up to the later survey waves, pooled. The only specification change is to control for survey fixed effects. Because the estimates of the return to schooling demand the most variables—both schooling and wages—they have the smallest samples, consisting of SUSENAS 2011–14 data for wage workers. Samples are 2–7 times larger than from 1995 alone.

The new schooling estimates are more sharply estimated than the old, yet do not differ statistically, which is expected since schooling attainment hardly changes during adulthood (compare Table 2 and Table 5). Still, there is a greater suggestion of positive impact. When sticking to the Duflo (2001) specification choices—no weights, inclusion of population-based controls—the original estimate of 0.12 extra years of schooling per unit of treatment (with clustered standard error 0.042) becomes 0.10 (0.032). The point estimate of the placebo effect rises from 0.015 (0.042) to 0.050 (0.041). The kink estimate—now more preferred because of this increased pre-trend—is consistently positive, unlike in the 1995 data alone (see Figure 7). Once more, the kink is not so significant without the child population controls, at 0.038–0.43 (0.038); still, these result provide modestly more evidence that Inpres SD boosted schooling attainment.

As for employment, the shift to prime working years turns out to introduce a downward pre-trend, especially when weighting. And low $p$ values on the Hausman test tends to justify weighting. See the left halves of Table 6 and Figure 8. Here too, the pre-trends increase the importance of the kink-based estimator. The weighted kink fit implies an impact of 1.6 (0.7) percentage points per unit of treatment with the child population controls, and 0.93 (0.62) without.

The treatment-wage association is smaller in the later follow-ups than in 1995, but still consistently positive. The weighted, kink-based log wage impact falls from 0.065 (0.026) in 1995 to 0.034 (0.022) in 2011–14 with the extra controls and from 0.057 (0.024) to 0.0088 (0.020) without (comparing the rightmost columns of Table 3 to Table 6 and Figure 4 and Figure 8).

The 2SLS results remain mostly weakly identified and ambiguous. For instance, the most sharply estimated wage impact estimate—from a preferred specification that includes weights and the kink model, but not the extra controls—returns a 95% confidence interval of [−0.24, 0.14].

Appendix E repeats the exercise of this section with the 1995 data included as well. Since samples are still dominated by the new, larger follow-ups the results remain similar. The 2SLS results are not well identified except in the specification closest to the original—without weights, with the extra controls, instrumenting by young/old (Table E–3, panel B). There, the point estimate for the return to schooling is 0.23 with a 95% confidence interval of [0.10, 0.67].

In sum, while returns to schooling are impossible to measure with any precision in the sample from prime earnings years, kink-based estimates of the impacts on schooling and (reduced-form) on wages are consistently positive, and with statistical strength when including a set of controls present in the original paper. Since the controls may have been chosen in light of their impact on the wage results, estimates including them deserve some discount, but not a complete discount.

**Table 5. Reduced-form TWFE impact estimates for years of schooling, pooled data for 2005, 2011–2014, 2017–19**

| | (1) | (2) | Hausman $p$: weighted vs. not |
|---|---|---|---|
| *A. With controls based on number of children aged 5–14* | | | |
| Experiment | 0.10 | 0.099 | 0.86 |
| | (0.032) | (0.036) | |
| Placebo | 0.050 | 0.10 | 0.053 |
| | (0.041) | (0.046) | |
| Kink | 0.086 | 0.069 | 0.55 |
| | (0.040) | (0.041) | |
| *B. Without extra controls* | | | |
| Experiment | 0.076 | 0.079 | 0.88 |
| | (0.031) | (0.035) | |
| Placebo | 0.054 | 0.094 | 0.12 |
| | (0.038) | (0.043) | |
| Kink | 0.043 | 0.038 | 0.85 |
| | (0.038) | (0.038) | |
| Data corrections | ✓ | ✓ | |
| Clustered | ✓ | ✓ | |
| Weighted | | ✓ | |
| Experiment $N$ | 589,671 | 589,671 | |
| Placebo $N$ | 486,230 | 486,230 | |
| Kink $N$ | 1,023,726 | 1,023,726 | |

*Notes:* Notes to Table 2 apply.

**Table 6. Reduced-form TWFE impact estimates for labor outcomes, 2005, 2011–14**

| | Formal sector employment | | | Log hourly wage | | |
|---|---|---|---|---|---|---|
| | (1) | (2) | Hausman $p$: weighted vs. not | (3) | (4) | Hausman $p$: weighted vs. not |
| *A. With controls based on number of children aged 5–14* | | | | | | |
| Experiment | 0.0020 | –0.0043 | 0.0032 | 0.022 | 0.0091 | 0.056 |
| | (0.0026) | (0.0032) | | (0.0088) | (0.011) | |
| Placebo | –0.0056 | –0.025 | 0.0000035 | –0.034 | –0.024 | 0.49 |
| | (0.0050) | (0.0068) | | (0.014) | (0.022) | |
| Kink | 0.0081 | 0.016 | 0.074 | 0.034 | 0.034 | 0.99 |
| | (0.0055) | (0.0067) | | (0.015) | (0.022) | |
| *B. Without extra controls* | | | | | | |
| Experiment | –0.0026 | –0.0077 | 0.014 | –0.0061 | –0.018 | 0.079 |
| | (0.0025) | (0.0029) | | (0.0087) | (0.0099) | |
| Placebo | –0.0062 | –0.021 | 0.00014 | –0.025 | –0.015 | 0.46 |
| | (0.0047) | (0.0065) | | (0.013) | (0.018) | |
| Kink | 0.0047 | 0.0093 | 0.25 | 0.013 | 0.0088 | 0.71 |
| | (0.0051) | (0.0062) | | (0.014) | (0.020) | |
| Data corrections | ✓ | ✓ | | ✓ | ✓ | |
| Clustered | ✓ | ✓ | | ✓ | ✓ | |
| Weighted | | ✓ | | | ✓ | |
| Experiment $N$ | 382,383 | 382,383 | | 113,738 | 113,738 | |
| Placebo $N$ | 325,963 | 325,963 | | 68,938 | 68,938 | |
| Kink $N$ | 667,593 | 667,593 | | 184,723 | 184,723 | |

*Notes:* See notes to Table 3. The 2005 data covers employment but not wages.

**Table 7. 2SLS estimates of the effect of years of schooling on labor market outcomes, pooled data for 2011–14**

| | Employment | | | | Log hourly wage | | | |
|---|---|---|---|---|---|---|---|---|
| | Instrument by young/old | | Kink instrument | | Instrument by young/old | | Kink instrument | |
| | (1) | (2) | (3) | (4) | (6) | (7) | (8) | (9) |
| *A. With controls based on number of children aged 5–14* | | | | | | | | |
| Coefficient | 0.023 | –0.055 | 0.13 | 3.12 | 0.31 | 0.32 | –0.44 | –0.11 |
| | (0.028) | (0.052) | (0.13) | (34.2) | (0.15) | (0.55) | (0.71) | (0.13) |
| | [−0.06, 0.11] | (−∞, 0.03] | (−∞, ∞) | (−∞, ∞) | [0.11, ∞) | (−∞, ∞) | (−∞, −0.01] ∪ [0.29, ∞) | (−∞, 0.06] |
| KP $F$ | 7.15 | 4.33 | 1.63 | 0.0082 | 2.56 | 0.22 | 0.56 | 3.72 |
| *B. Without extra controls* | | | | | | | | |
| Coefficient | –0.043 | –0.11 | 0.33 | –0.27 | 0.25 | 0.27 | –0.024 | 0.022 |
| | (0.051) | (0.077) | (1.50) | (0.69) | (0.35) | (0.18) | (0.085) | (0.061) |
| | (−∞, 0.03] ∪ [1.03, ∞) | (−∞, −0.02] | (−∞, ∞) | (−∞, ∞) | (−∞, ∞) | (−∞, ∞) | (−∞, 0.11] | [−0.24, 0.14] |
| KP $F$ | 3.70 | 3.75 | 0.051 | 0.23 | 0.30 | 1.45 | 4.01 | 8.37 |
| Data corrections | ✓ | ✓ | ✓ | ✓ | ✓ | ✓ | ✓ | ✓ |
| Clustered | ✓ | ✓ | ✓ | ✓ | ✓ | ✓ | ✓ | ✓ |
| Weighted | | ✓ | | ✓ | | ✓ | | ✓ |
| Pre-trend control | | | ✓ | ✓ | | | ✓ | ✓ |
| Observations | 382,383 | 382,383 | 606,781 | 606,781 | 113,738 | 113,738 | 169,020 | 169,020 |

*Notes:* See notes to Table 4 and Table 5.

**Figure 7. Coefficients on interactions between age dummies and Inpres intensity in schooling regressions, 2005, 2011–14, 2017–19**

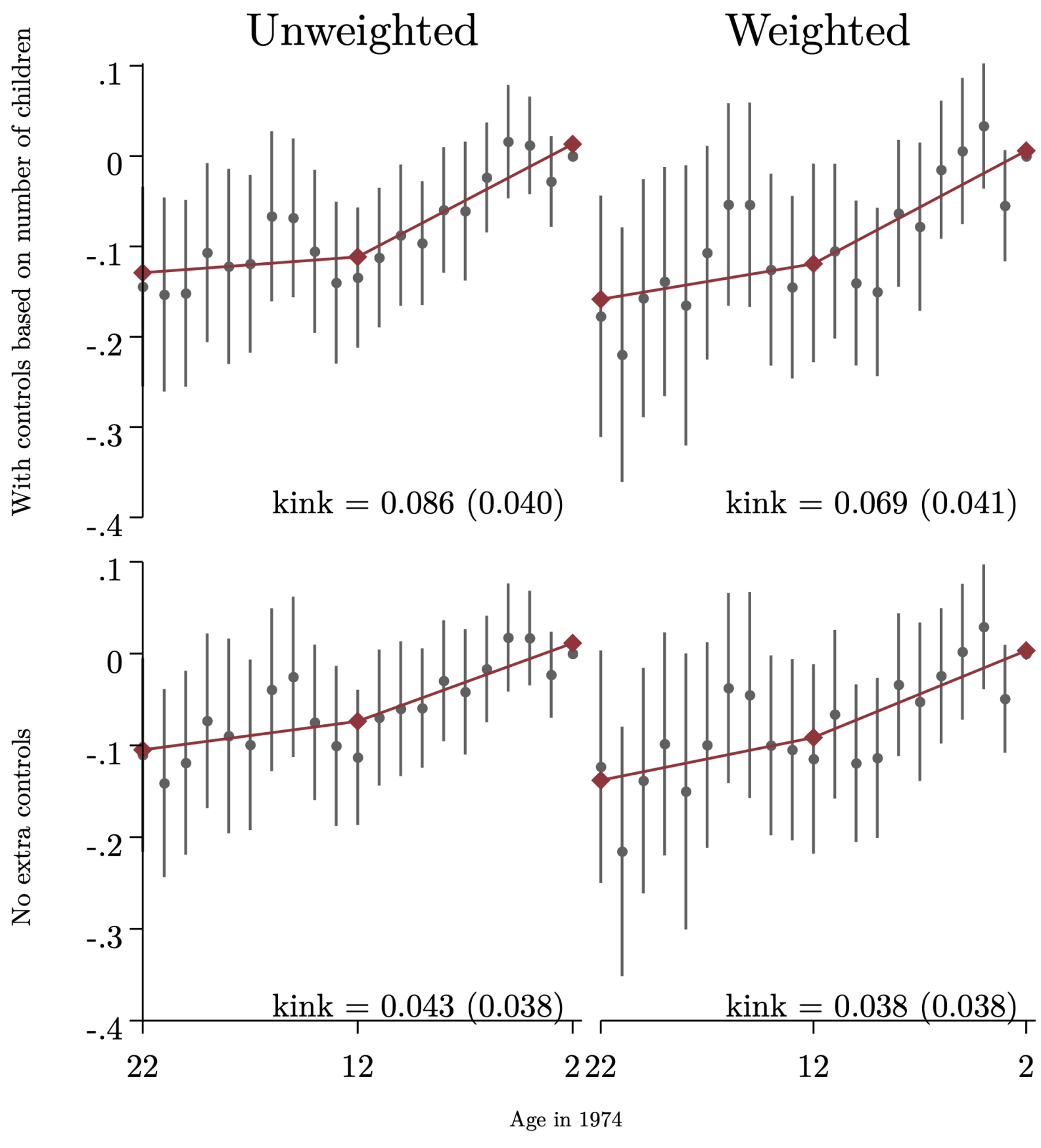

*Notes:* See notes to Figure 3.

**Figure 8. Coefficients on interactions between age dummies and Inpres intensity in labor outcome regressions, 2005, 2011–14**

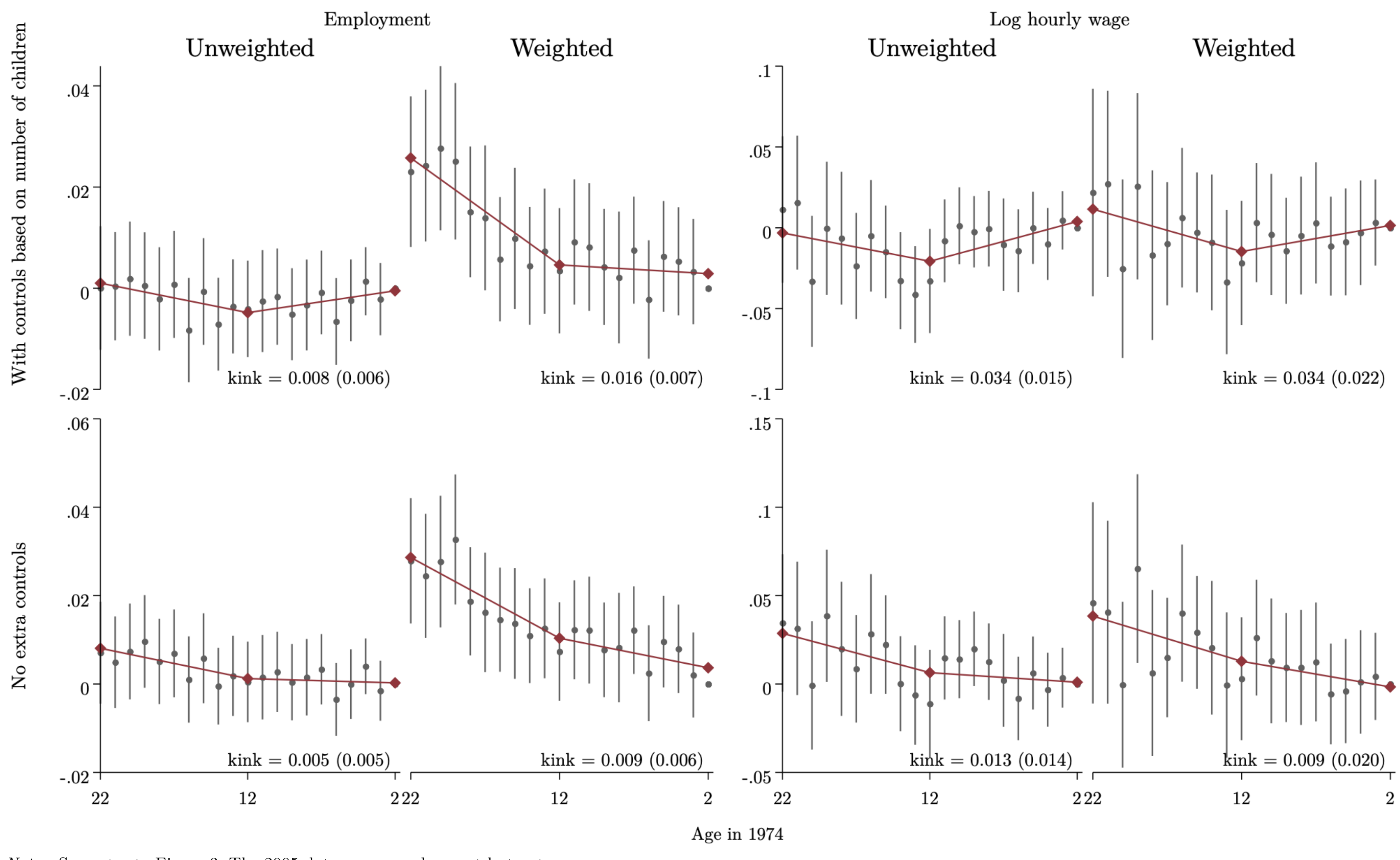

*Notes:* See notes to Figure 3. The 2005 data cover employment but not wages.

**Figure 9. Confidence curves from wild-bootstrapped tests of the coefficient on schooling in 2SLS regressions in Table 7, 2005, 2011–14**

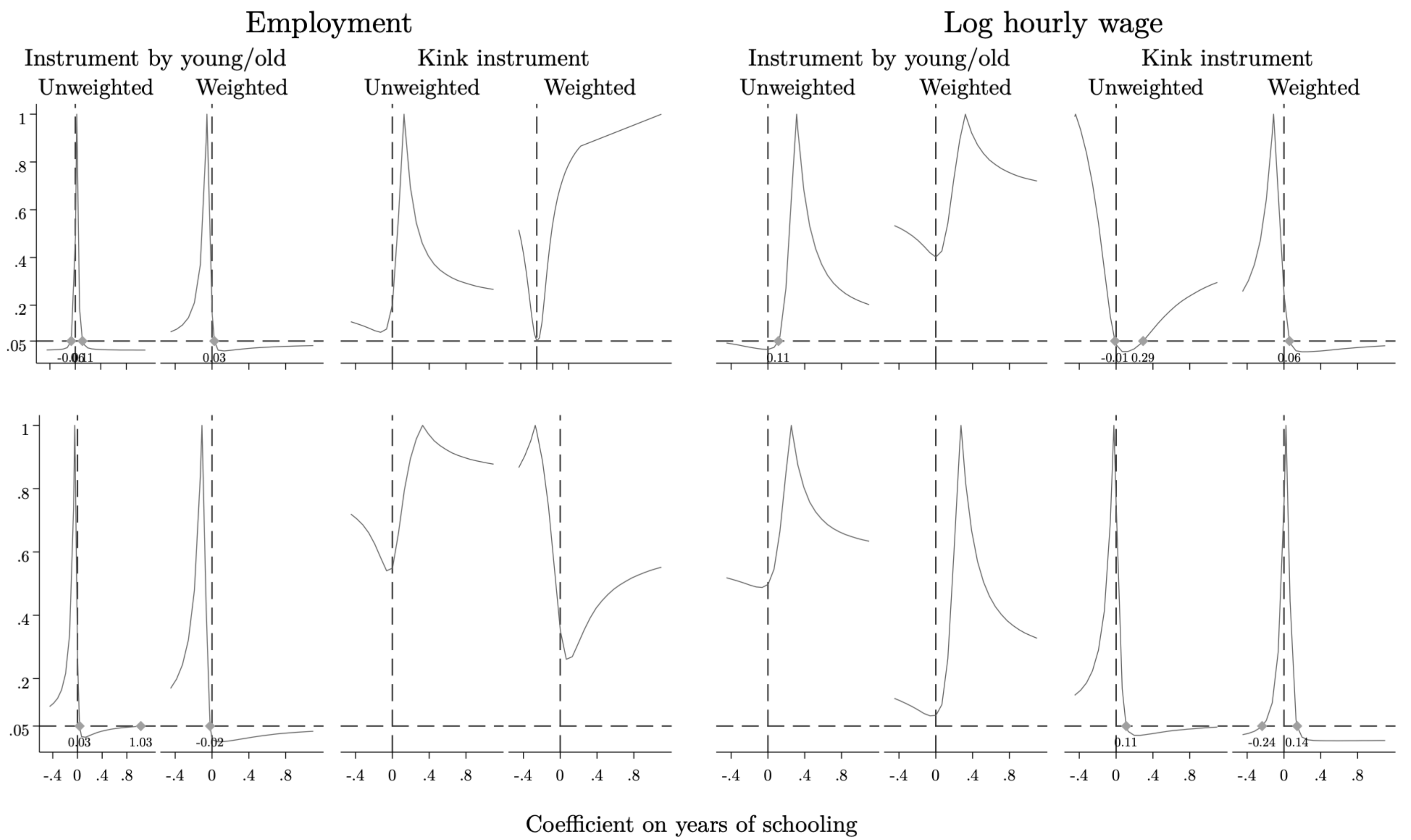

*Notes:* See notes to Figure 5 and Table 7.

## 5 Impacts by grade level

Recall from section 1 that Duflo (2001) disaggregates the main specifications along two dimensions to check for fingerprints of Inpres SD. The analysis so far has heavily emphasized the kink model, which is rooted in one of those checks. Along the way, it has raised some doubt about whether Inpres caused substantial breaks in the relative schooling trends at age 12 as of 1974. This section turns to the other fingerprint test, for the grade-by-grade impacts of the school construction impulse, which Duflo (2001) calls DID-in-CDF analysis. The examination will corroborate the view that Inpres affected primary schooling patterns, but in a surprisingly circumscribed way. Inpres seems to have increased primary school completion, but not so much total *years* spent in primary school. A potential confounder will also be exposed that further contributes to the seeming muted impact on total schooling.

This fingerprint check consists in running a set of regressions like (2)—meaning without any controls based on child population—but with the treatment variable dichotomized as in 2×2 DID. The dependent variables are dummies for whether a person has at least 1 years of schooling, or 2, or 3, and so on. The impacts estimates for all these outcome indicators are plotted, along with 95% confidence intervals.

Duflo (2001, Figure 2) is almost exactly replicated in the upper left of Figure 10, below.[27] The impact of Inpres on schooling attainment builds through the six years of primary school. This effect mostly evaporates in the three years of lower secondary school (grades 7–9). These results are consistent with Inpres mainly affecting primary school careers. Then, in upper secondary school (grades 10–12) as well as the early tertiary years, the "impact" goes negative. "The negative difference in differences at the senior high school level," Duflo (2001) reasonably surmises, "may indicate that some variable predicting the probability of attending senior high school is omitted from this regression (and changed in low program regions more than in high program regions)."

I vary this replication in the same ways as in previous sections. In the second row, I make the treatment variable continuous (planned schools per 1,000 children), incorporate data corrections, cluster standard errors, weight observations, and incorporate all feasible surveys (1995, 2005, 2011–14, 2017–19). Other than a modest shift toward zero in the lower secondary years, the pattern of results hardly changes.

Then, in the right column, I transfer these specifications to the placebo experiment (comparing ages 12–17 as of 1974 to 18–24). As before, I scale these estimates by 10.5/6.5 for comparability. Notably, and especially in the larger sample represented in the second row, the experiment and placebo patterns are similar. This suggests that most of the grade-wise fingerprint detected in Duflo (2001) is characteristic of long-term convergence, not Inpres specifically. However, some of the placebo effects in primary school are smaller, leaving open the possibility—to which we will return—that Inpres *accelerated* convergence. (Although the vertical axes are left unmarked for legibility, they are the same within each row of the figure.) Meanwhile, the negative "impact" on secondary schooling largely disappears in the placebo experiment. The omitted variable detected by Duflo (2001) apparently only affects the experiment.

[27] The code Duflo sent me does not produce this figure.

What might the variable be? A good candidate is Indonesia's contemporaneous expansion in *secondary* schooling. Under President Suharto, Indonesia managed development policy through five-year plans (Repelitas). One reason the government so quickly translated the oil windfall of late 1973 into the Inpres SD program is that Repelita II was near finalization. Much more so than Repelita I, it prioritized schooling. Indeed, it set targets: by 1979, 85% of primary-age children would be in school; 85% of primary school graduates would continue to middle school; and 81% of middle school graduates would continue to high school (Heneveld 1978, p. 63). Heneveld (1979, p. 147) describes the scale of state investment in secondary education as "staggering": nearly 1,000 new schools built, another 2,000 rehabilitated, more than 25,000 teachers hired. The growth continued under Repelita III. The annual number of middle school graduates doubled between 1974 and 1979, from 293,000 to 595,000, then doubled again by 1984, to 1.25 million. High school graduating classes mushroomed too, from 78,000 to 154,000 to 456,000 (IEES 1986, pp. 6, 8). While the number of children in primary school rose 32% between 1973 and 1977, the number in secondary school grew twice as fast, 66%, from a smaller base. In contrast with primary school students, the majority of new secondary students went to *private* schools, which indicates that the supply expansion, however great, was outstripped by demand. And the demand was not evenly distributed. "The most rapid increases in demand have occurred precisely in those areas which are profiting most from Indonesia's new economic growth, the cities, and in East Java" (Heneveld 1978, p. 74). Since Javanese regencies had higher populations, they tended to receive fewer Inpres schools per child. The intensity of the primary and secondary "big pushes" may have been anti-correlated across regencies.

Likely, then, the effects of the Inpres SD primary school expansion on *total* schooling are partially confounded by a secondary schooling expansion that disproportionately affected the regencies that got less Inpres SD treatment. This might help explain why the impact on *total* schooling was earlier found to be somewhat weak.

To sharpen the picture of the schooling impacts, the third row of Figure 10 introduces another methodological change: each regression for attending a grade is restricted to completers of the previous grade, making the dependent variables dummies for *continuation*. Since the samples are small for the highest schooling levels, their confidence intervals are wide and are dropped to avoid distorting the vertical scale. In the primary years, we continue to see positive impacts at each grade. The largest seeming impact is in grade 1—that is, in making it practical for children to attend elementary school at all. However, the same effect is seen in the placebo group. Meanwhile, back in the experimental comparison, the negative results in the secondary grades now concentrate in grades 7 and 10, i.e., on whether students continue to the next schooling level.

To systematize the experiment-placebo comparisons, the bottom row of the figure brings the kink model to these finely sliced schooling variables, once more allowing a trend break at age 12 as of 1974. The plot of kink estimates in the bottom left of Figure 10 shows no clear trend break for continuation into grade 7 but does show large negative kinks at grades 10 and 13. These imply that the relative slow-down in continuation into lower secondary school in high-Inpres areas did not break with the historical trends, while the slow-downs in continuation into upper secondary and university did.

Somewhat surprisingly, within the primary school years, by far the largest deviation from historical trend was in continuation into 6$^{th}$ grade, the final year of primary school. One reason might be an inappropriate choice of kink point. The reason to expect a kink in grade 6 continuation at age 12 as of 1974 is the same as for total schooling: the 11 year-olds were the first cohort to be able to attend the new schools—which they would only have been able to do for one year, in grade 6. By the same reasoning, we might expect a kink in continuation into grade 5 to appear one cohort later, and so on down to a kink in grade 1 continuation at the cohort aged 7 in 1974. Figure 11 investigates—and mostly rejects—this theory. The kink plots for continuation at each grade level are constructed in the usual way except that the spline is shifted in the way just described. The kink in grade 6 continuation remains far larger than the others, which are themselves diverse in sign.

In sum, as far as the data go, the trend break in schooling attributable to Inpres SD is largely confined to continuation into the final year of primary school. This effect may real, or it may reflect measurement noise in the lower grades. Perhaps survey respondents accurately recall whether they completed primary school but, conditional on not completing it, do not accurately recall when they stopped.

The concentration of apparent continuation in the final grade and the contrary developments at the secondary and even tertiary levels may both mask the impact of Inpres on schooling careers in the main Duflo (2001) regressions. If so, using a narrowed schooling variable might improve identification. There are two natural candidates: total *primary* schooling, capped at 6 years; and a dummy for primary completion, more precisely for whether a person attended grade 6 or higher. Figure 12 studies the impact of Inpres on these variables using kink plots. Total primary schooling shows remarkably little trend break upon the arrival of Inpres—another surprise—but primary completion does, making the latter a more promising candidate for kink-based 2SLS estimation of the returns to schooling. Appendix F reports 2SLS estimates of the impact of primary completion on employment outcomes in 2011–14. Once more, the estimates are imprecise.

Bringing the specification changes developed in previous sections to the Duflo (2001) DID-in-CDF analysis reveals a sort of two-dimensional fingerprint of Inpres, spread along the dimensions of schooling grade and birth year. The results support the view that Inpres SD accelerated convergence in schooling between low- and high-treatment areas, at least in primary completion. Also exposed is the likely fingerprint of simultaneous post-primary schooling expansion. Unfortunately, removing this confounding variation does not sharpen the estimates of returns to schooling in the Inpres quasi-experiment.

**Figure 10. TWFE estimates for schooling attainment and continuation**

Experiment | Placebo

Original

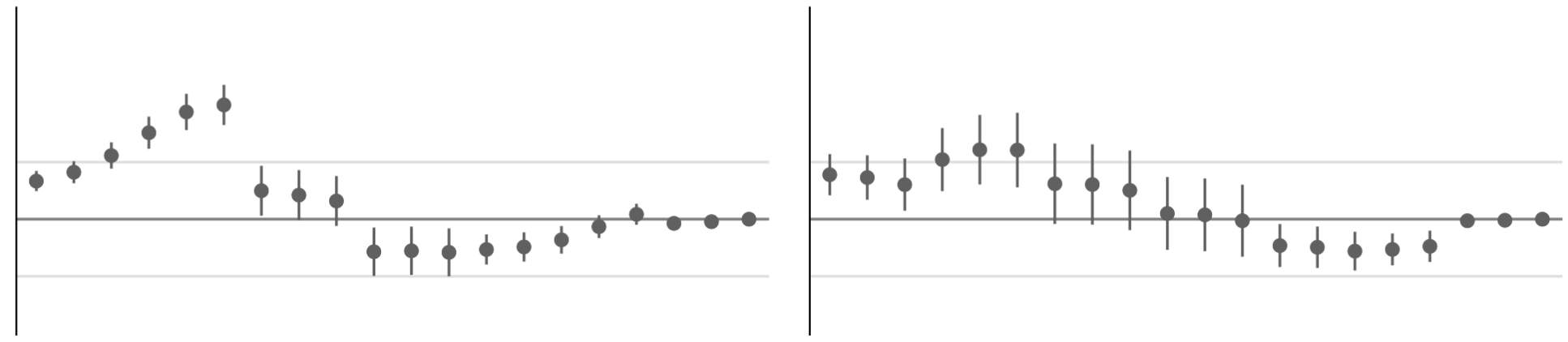

+ corrected, clustered, weighted, continuous treatment, all years with data

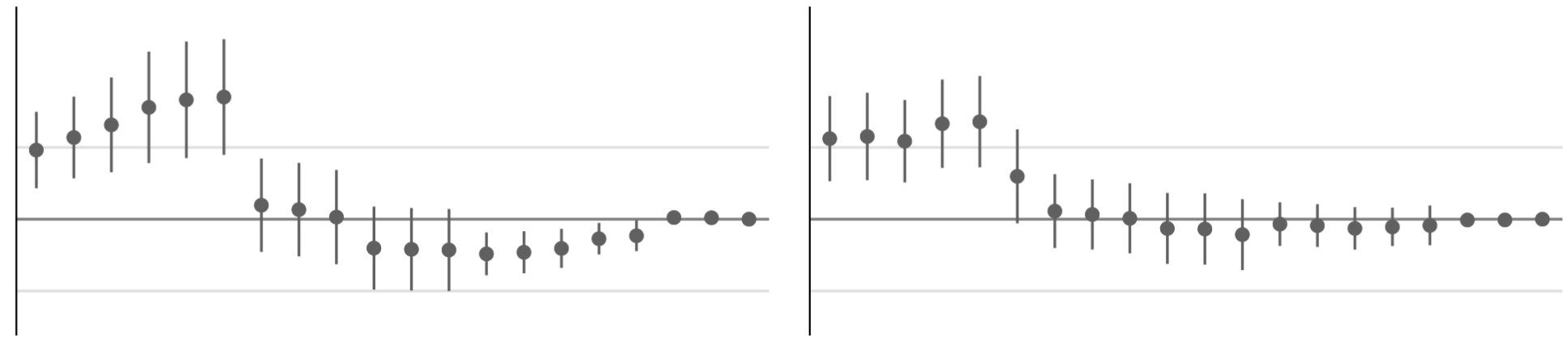

+ conditional on attending previous grade

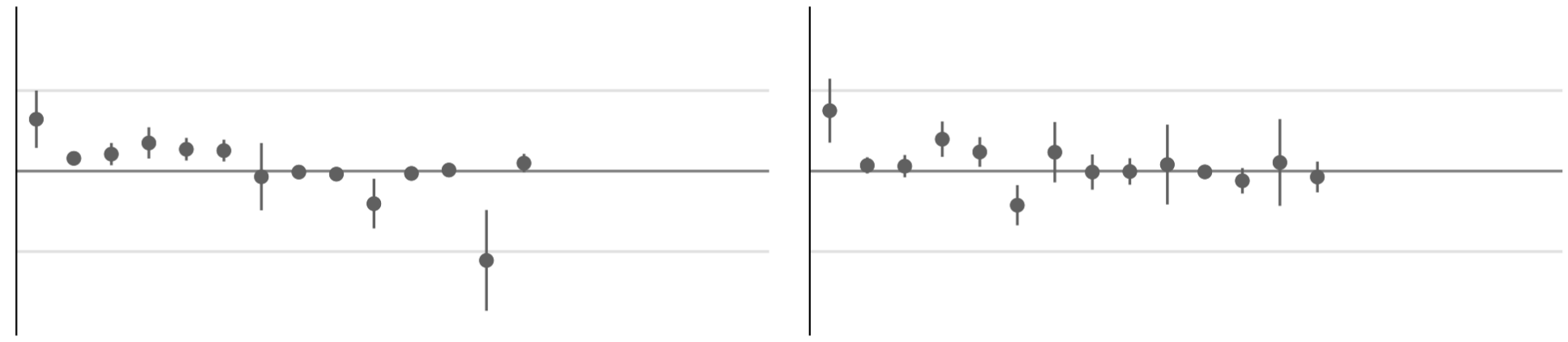

+ kink model

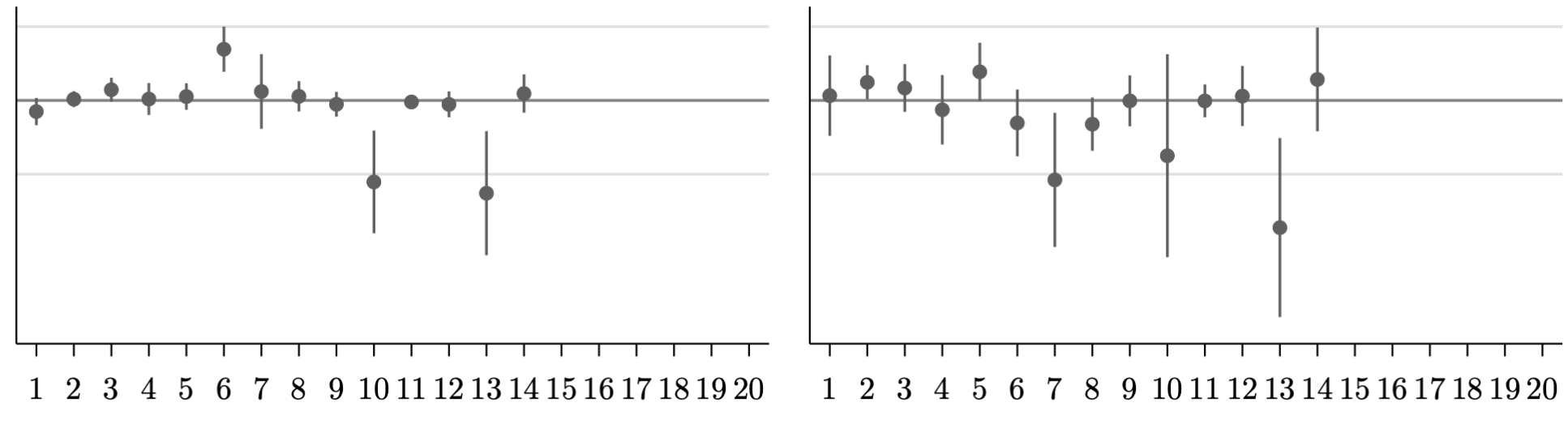

Years of schooling

Notes: Though not marked, the vertical axis ranges are the same within each row. Each point estimate and 95% confidence interval is from a separate TWFE regression, incorporating fixed effects for year and regency of birth, No other controls are added, except the one included by definition in the kink model. The dependent variables are dummies for whether a person attend school for $\geq m$ years, $m = 1, \ldots, 20$. The upper left plot is a close replication of Duflo (2001), Figure 2.

**Figure 11. Kink estimates for schooling continuation rates, all available survey rounds**

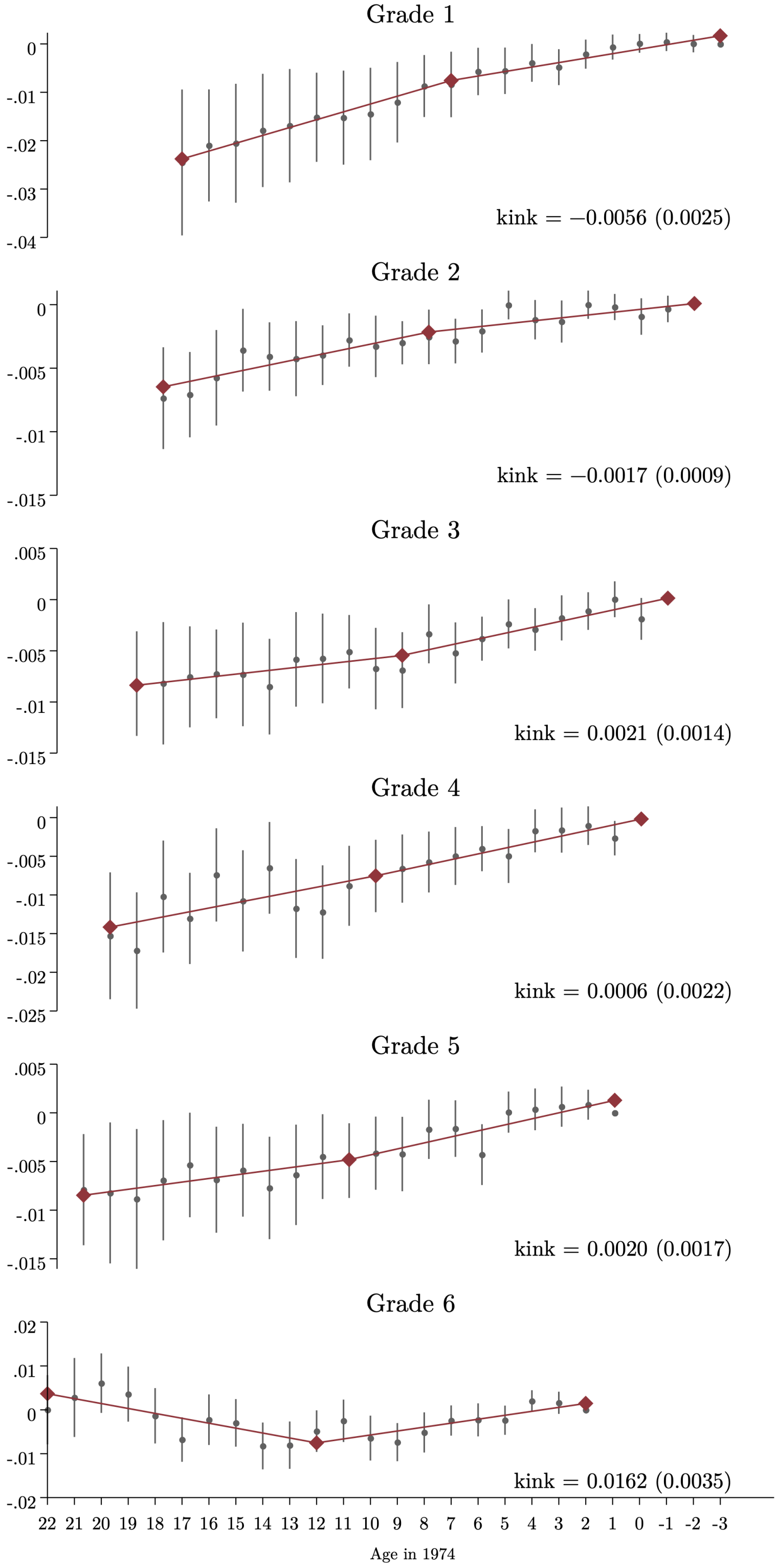

Notes: The plots are constructed in the same way as in earlier figures except that the spline is shifted as shown. The dependent variable is attendance of a given grade conditional on having attended the previous grade. No control based on child population are included.

**Figure 12. Kink estimates for primary schooling and primary completion, all available survey rounds**

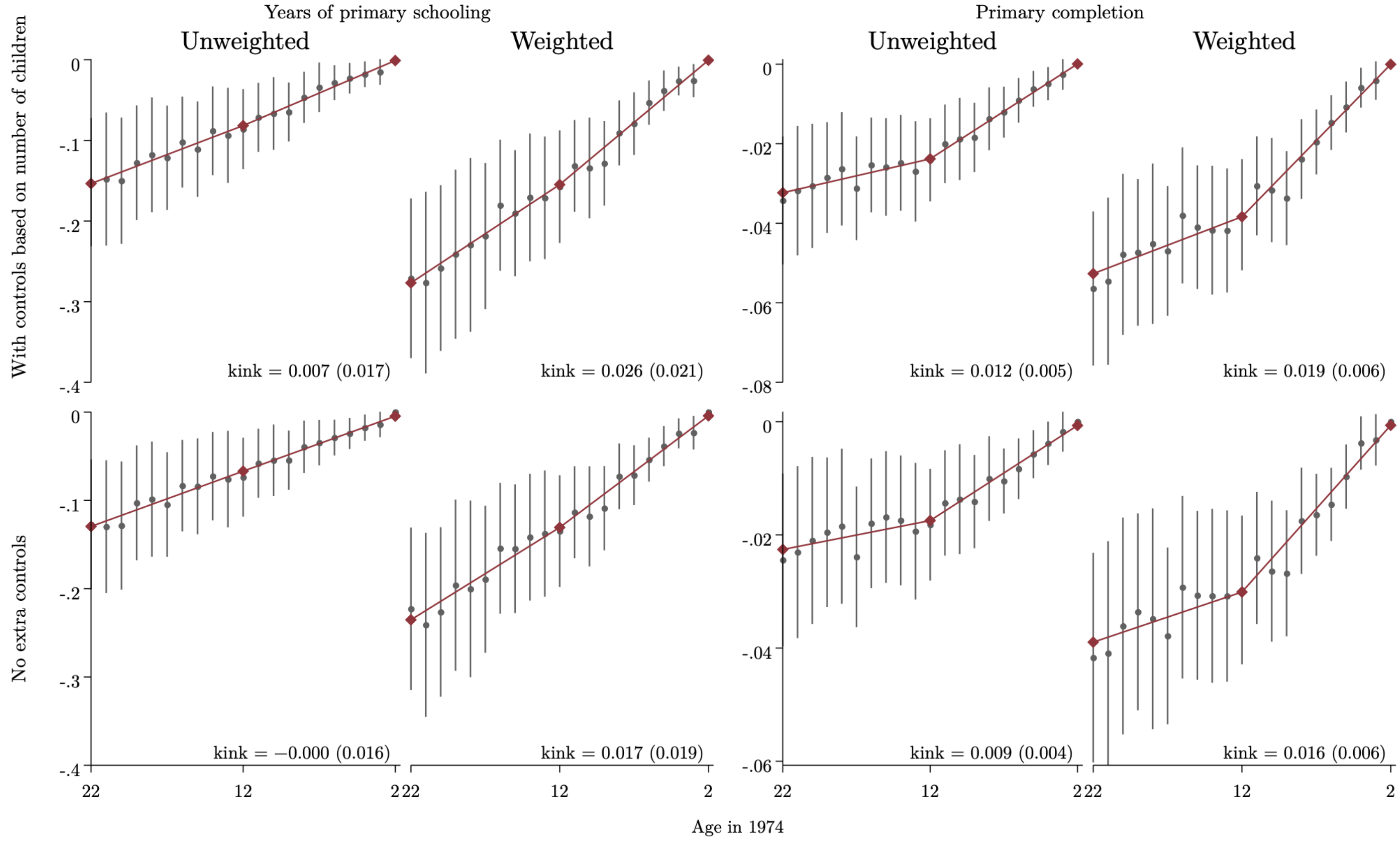

Notes: See notes to Figure 3. Data are for 1995, 2005, 2011–14, 2017–19. No control based on child population are included.

## 6 Changes in changes

An alternative to controlling for nonparallel trends, as with the kink model, is to apply an estimator that is intrinsically robust to their presence. A leading example is the quantile-based changes-in-changes (CIC) estimator of Athey and Imbens (2006). Appendix G reports results from applying this estimator to reduced-form wage impacts. Despite the looser identifying assumptions, the standard errors are similar to those from the kink-based estimates in Table 3 and Table 6. The strongest finding is of positive wages impacts at low deciles during the prime earnings years (2011–14). The progressive impact (in the distributional sense) is consistent with causation by Inpres, which would primarily have lengthened the schooling careers of poorer people, for whom the marginal year was more often in primary school. There is also a hint of negative impacts at high deciles, which might reflect the confounding from post-primary schooling expansions. Appendix H explains why the "Wald CIC" estimator of de Chaisemartin and D'Haultfœuille (2017) is not deployed to estimate the returns to schooling.

## 7 Conclusion

The scrutiny here brought to Duflo (2001) is a testament to its enduring importance. Few studies so credibly claim to identify the impacts of schooling at scale. Most of the comments assess the robustness of the results to relaxing certain identifying assumptions. And many stem from the passage of time, which has brought more data and econometric insight. Several potential sources of bias in estimates and standard errors are pointed out. Steps to address these concerns, including weighting and spline-based modeling, tend to reduce power; and one should not equate the failure of a lower-powered test to reject the null with proof of the null. Still, the addition of data from later follow-ups partly compensates for the loss of power.

How is the evidence best read? Four changes introduced here are relatively uncontroversial: correcting baseline data, clustering standard errors by geographic treatment unit, following up later, and employing weak identification–robust inference. Others are best seen as matters of judgment, making tradeoffs that are typically clear in principle and hard to measure in practice. In various combinations, the debatable changes produce an ambiguous set of results, which frustrate the desire for a clean narrative. Since many of the estimates are insignificant at 0.05, one might declare the central findings of Duflo (2001) fragile. However, such a binary judgment would destroy information. The Duflo (2001) "fingerprint" evidence partly persists, and is hard to explain as pure noise. The detected trend breaks in schooling, employment, and wages are almost all positive. Decomposing the schooling impacts by grade exposes a distinctive impact on primary school completion. My reading is therefore that the natural causal story, from school construction to schooling attainment to employment wages *does* best explain the results. It is reasonable to come to the study with a prior that puts weight on this story, and the evidence reported here will reinforce that prior. Still, while I think that is the best explanation, my certainty is low. The same evidence would hardly persuade a reader who arrived much more skeptical. And the returns to schooling appear almost impossible to measure with any precision.

I hope that researchers will continue to take inspiration from Duflo (2001) as a creative, rigorous effort to extract evidence from a natural experiment that is relevant to major theoretical questions and global issues. That said, this reanalysis dramatizes that the hunt for causal truth outside of actual experiments is at least as hard as valuable.

## Appendices

# A. Testing for effect heterogeneity

Recent scholarship (de Chaisemartin and D'Haultfœuille 2020; Goodman-Bacon 2021) has highlighted how effect heterogeneity can bias TWFE estimates of mean effects. A key insight is that controlling for the two sets of fixed effects rearranges as well as restricts the identifying variation in the treatment. For example, if treatment is binary, then treated observations may effectively be assigned negative treatment after partialling out the fixed effects, while untreated observations may effectively be assigned positive treatment. When combined with certain patterns of effect heterogeneity, these shifts can reverse the sign of the effect estimate.

In this context, the Jakiela (2021) test for homogeneity is straightforward. One partials the fixed-effect dummies out of the treatment and outcome variables. Then one regresses the "residualized" outcome on the residualized treatment—separately for the treated and untreated subsamples. If the treatment effect is homogeneous, the two slope estimates will not differ (though the converse is not true).

I run the test on a preferred variant of a Duflo (2001) regression. In the penultimate columns of Table 2 and Table 3 appear results from reduced-form regressions of years of schooling and the log hourly wage in 1995 on the treatment intensity indicator $D_j T_t$. The samples are restricted to men aged 2–6 and 12–17 in 1974. The minimal Duflo (2001) control set is included. Data corrections are incorporated. Standard errors are clustered by birth regency. Observations are weighted.

I modify the Jakiela test in two minor ways. Since treatment is not binary, I split the sample according to the (corrected) Duflo (2001) indicator of high-versus-low treatment. And since the representative regression includes other controls, I partial them out too. Figure A–1 depicts the data so processed. The blue and black circles are for treated and untreated observations respectively. For clarity, values are grouped and averaged for each age cohort–birth regency pair. Superimposed on the two scatter plots are linear and local polynomial smoothed fits. For schooling and the log hourly wage, the slopes of the linear fits are nearly identical. The $p$ values for the nulls that they match are 0.61 and 0.91.

This check for treatment heterogeneity therefore finds essentially none.

**Figure A–1. Residualized outcome vs. residualized treatment in revised Duflo (2001) reduce-form wage regression, age and birth regency**

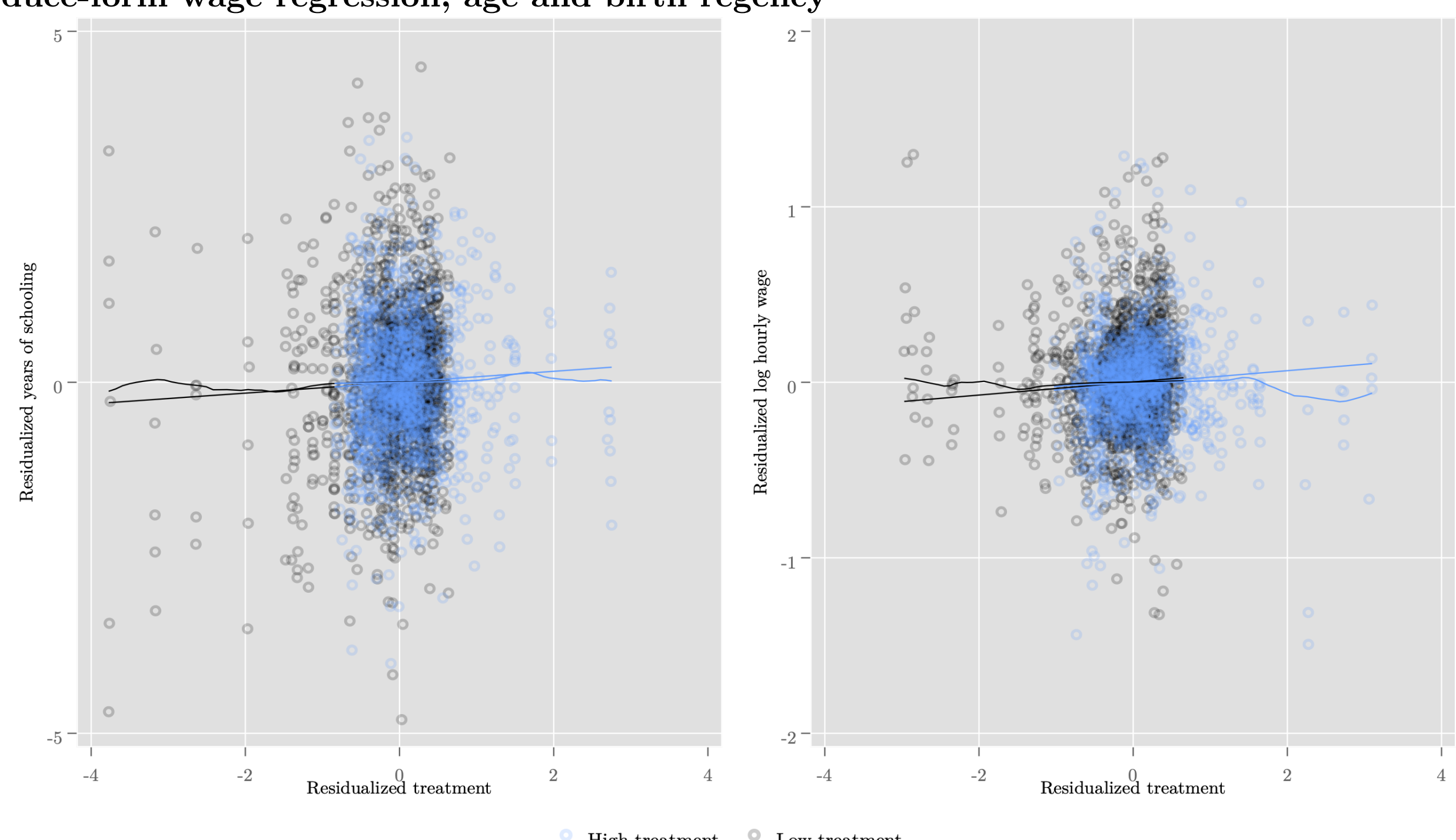

## B. Weighting, consistency, and efficiency

Section 2.6 points out that if in the context of a particular estimation problem, sampling is endogenous, then weighting observations in inverse proportion to their sampling probability may be necessary for consistency. Typically, though, it increases the variance of estimators. This appendix investigates more precisely how the sampling program affects the consistency of OLS and its efficiency relative to weighted OLS (WOLS).

Consider the homogeneous treatment effect model

$$y = \alpha + x\beta + e \tag{5}$$

Here, $x$, $e$, and $y$ are variables in the population, with realizations $(x_i, e_i, y_i)$ that are sampled with independent probabilities $p_i$. For the moment, the standard exogeneity assumption $\mathrm{E}[e|x] = 0$ is not imposed to identify $\alpha$ and $\beta$. Instead, $\alpha$ and $\beta$ are *defined* as the results from (unweighted) OLS regression of $y$ on $x$ in the population.[28] This implies only that $\mathrm{E}[e] = 0$.

Suppose we perform OLS of $y$ on $x$ within the sample. What regression in the *population* will produce the same result in the large-population, large-sample limit? The answer: regression of $y$ on $x$ while weighting by $p$. While this population regression would weight each observation by its probability of observation, the unweighted sample regression would weight each observation in the population by its *realized* chance of observation, 0 or 1. Thus, asymptotically, the unweighted OLS estimate of $\beta$ in the sample is

$$\hat{\beta}_u = \frac{\mathrm{Cov}_p[x, y]}{\mathrm{Var}_p[x]}$$

where the variance is computed over the population and subscripts indicate weighting by $p$. Substituting with (5), the asymptotic bias of $\hat{\beta}_u$ is

$$\hat{\beta}_u - \beta = \frac{\mathrm{Cov}_p[x, \alpha + x\beta + e]}{\mathrm{Var}_p[x]} - \beta = \frac{\mathrm{Cov}_p[x, e]}{\mathrm{Var}_p[x]} \tag{6}$$

[28] This estimand is one considered in Brewer and Mellor (1973) and DuMouchel and Duncan (1983).

$\hat{\beta}_u$ is unbiased for $\beta$ if and only if the numerator on the right is 0.

There are at least two circumstances under which this condition holds:

1. *The sampling probability is exogenous.* This condition permits $p$ to be correlated with $x$, but not with $e$: $\mathrm{Cov}[p, e] = 0$. In fact, it also requires $\mathrm{Cov}[px, e] = 0$. I will call these two requirements second- and third-order conditions since they involve a second and third moment of the variables.[29]
2. *The sampling probability is purely endogenous.* The second-order condition here is the mirror image of the previous one. $p$ may be correlated with $e$ but not $x$: $\mathrm{Cov}[p, x] = 0$. Here too, a less-intuitive third-order condition is also required: $\mathrm{Cov}[pe, x] = 0$.[30]

To bring these mathematical statements to life, I simulate several data generating and sampling processes, apply OLS and WOLS in each case, and summarize the results in Figure B–1. In all cases, a population is generated with 5,000 independent standard normal draws for both $x$ and a primary error term $\varepsilon$. The outcome is then computed as $y = x + e$ where $e$ is set to $\varepsilon$ (homoskedastic error; first row) or $\varepsilon(x + 3)/5$ (heteroskedastic; second row). For each observation, the sampling probability $p$ is 1 or 0.05 depending on whether $x > 0$ (purely exogenous; first column), or whether $x$ and $e$ have the same sign (second), or solely whether $e > 0$ (purely endogenous; third column), or $y > 0$ (fourth). Each pane shows the asymptotic best-fit line in the population, $y = x$, along with scatter plots of the samples, and OLS and WOLS fits thereto. Confidence intervals for the fits are also shown; for legibility they are some 11 standard errors wide, corresponding to the 99.999995% confidence level.

The first column of Figure B–1 confirms that under exogenous sampling, OLS is consistent—even in the presence of heteroskedasticity. In both panes of the column the OLS and WOLS best-fit lines all but coincide. Here, the sampling probability depends deterministically on $x$ alone, which suffices to satisfy both conditions for the consistency of OLS under exogenous sampling—that $p$ and $px$ are uncorrelated with $e$ in the population. Under homoskedasticity (top of column 1), OLS is also more efficient than WOLS, with its narrower confidence band rendered in purple. Under heteroskedasticity (next pane down), weighting by the inverse sampling probability happens in this case to counteract the inefficiency created by unequal error variances, by up-weighting the lower-variance observations on the left. (If $p$ were higher for $x < 0$ instead of lower, then the weighting would compound rather than counteract the inefficiency caused by the heteroskedasticity.) Being thinner, the green band for the weighted estimator is fully obscured by the purple one for the unweighted estimator. The backgrounds of both panels in the left column are slightly shaded to indicate that here, unweighted OLS is consistent. In sum, under exogenous sampling, OLS is consistent and typically more efficient than WOLS, though heteroskedasticity can augment or counteract this efficiency advantage.

[29] The condition is equivalent to $\mathrm{E}[pe] = \mathrm{E}[p]\,\mathrm{E}[e]$ and $\mathrm{E}[pxe] = \mathrm{E}[px]\,\mathrm{E}[e]$, both of which are 0 since $\mathrm{E}[e] = 0$. Define $\bar{p}$ as the normalized sampling probability $p/\mathrm{E}[p]$. Then the numerator on the right side of (6) is $\mathrm{Cov}_p[x, e] = \mathrm{Cov}_{\bar{p}}[x, e] = \mathrm{E}[\bar{p}xe] - \mathrm{E}[\bar{p}x]\,\mathrm{E}[\bar{p}e] = 0 - \mathrm{E}[\bar{p}x] \cdot 0 = 0$.

[30] The conditions now imply $\mathrm{E}[px] = \mathrm{E}[p]\,\mathrm{E}[x]$ and $\mathrm{E}[pex] = \mathrm{E}[pe]\,\mathrm{E}[x]$. The numerator in (6) can now be developed as $\mathrm{Cov}_p[x, e] = \mathrm{E}[\bar{p}xe] - \mathrm{E}[\bar{p}x]\,\mathrm{E}[\bar{p}e] = \mathrm{E}[\bar{p}e]\,\mathrm{E}[x] - \mathrm{E}[\bar{p}]\,\mathrm{E}[x]\,\mathrm{E}[\bar{p}e] = \mathrm{E}[\bar{p}e]\,\mathrm{E}[x]\,(1 - \mathrm{E}[\bar{p}])$, in which $\mathrm{E}[\bar{p}] = 1$.

The second column shows why the third-order condition for exogenous sampling is necessary. Here, sampling is dense when $x$ and $e$ have the same sign. We still have the naïve exogenous sampling condition, $\text{Cov}[p, e] = 0$, because knowing the value of either variable tells one nothing about the expectation of the other. For example, the average error is the same—zero—in the densely and lightly sampled regions. But $\text{Cov}[px, e] \neq 0$: where $p$ and $x$ are both at their largest in magnitude, in the lower left and upper right, the average of $e$ is too. So while WOLS remains consistent, OLS does not.

The third and fourth columns move to endogenous sampling. The third illustrates purely endogenous sampling as defined above: $p$ depends only on $e$. Under homoskedasticity (top row), WOLS is once more consistent, at least for the slope of the best fit. This time the counterexample demonstrating the need for the third-order condition appears in the same column, in the second row. It is possible for heteroskedasticity to cause $\text{Cov}[pe, x] \neq 0$. Here, when $p$ and $e$ are at their largest, so is $x$. This makes OLS inconsistent even for the slope.

The last column of Figure B–1 shows that if $p$ is related directly to the outcome $y$, as distinct from its unexplained component $e$, then unweighted OLS is inconsistent even under homoskedasticity. An exception here would be if the true best-fit line were horizontal—if there were no effect—for then direct dependence on $y$ would be equivalent to direct dependence on $e$, bringing us back to the third column.

WOLS is consistent throughout. Usually it has higher variance than OLS. But in these examples, when only WOLS is consistent, its mean squared error—variance plus the squared bias—is lower.

**Figure B–1. Illustration of unweighted and weighted regression in the presence of disproportionate sampling**

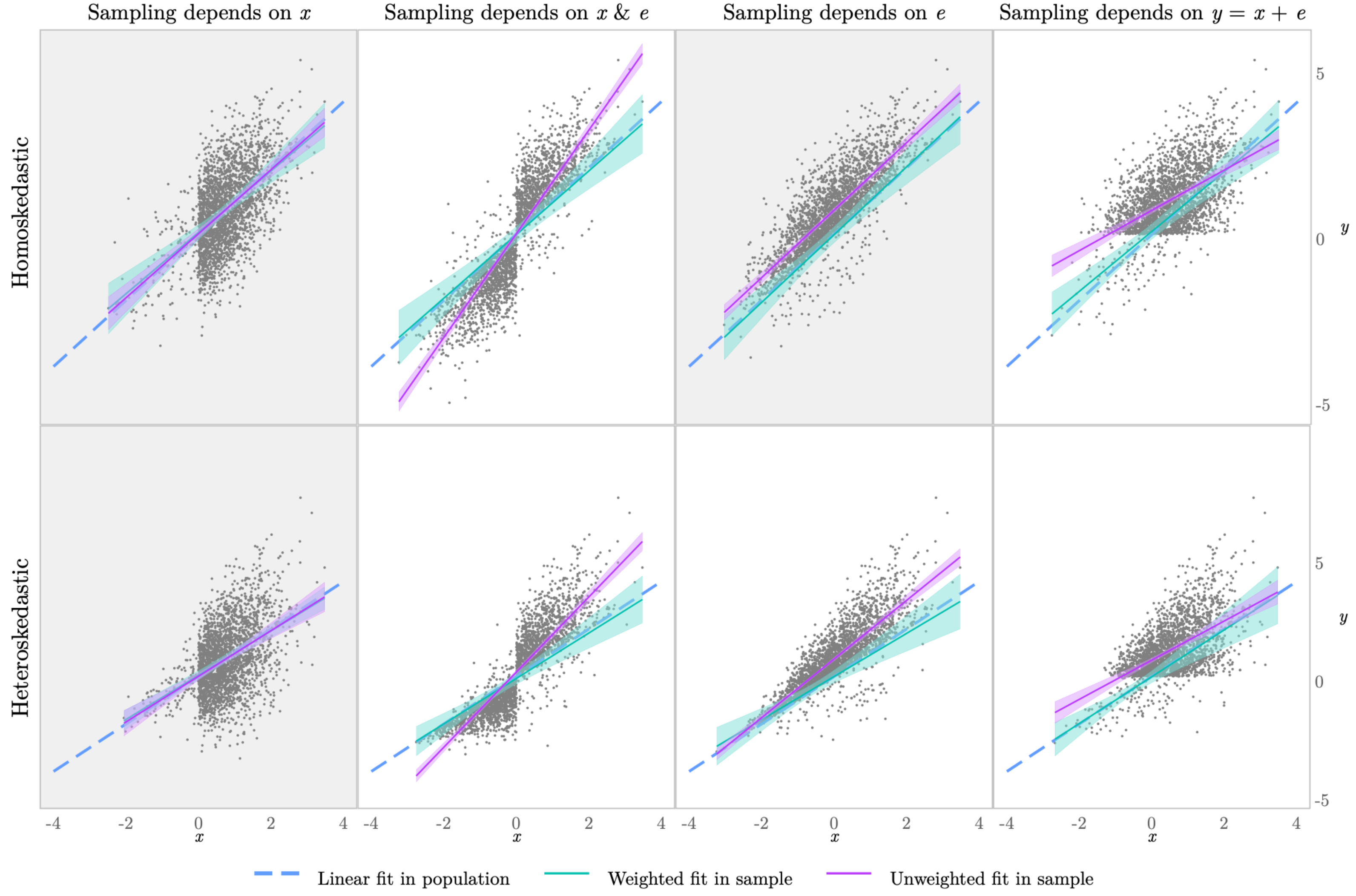

Notes: For all panels, a single population of 5,000 independent standard normal draws for $x$ and a primary error term $\varepsilon$ are drawn. The outcome is constructed as $y = x + e$ where $e$ is $\varepsilon$ (first row) or $\varepsilon(x + 3)/5$ (second row). The sampling probability is 1 or 0.05 depending on whether or not $x > 0$ (first column), or $x$ and $e$ have the same sign (second), or $e > 0$ (third), or $y > 0$ (fourth). The OLS fit in the population is plotted along with fits on the sample with and without weighting by $1/p$. For legibility the confidence level on the confidence intervals is 99.999995%. Confidence intervals are robust to heteroskedasticity except in the unweighted fits in the first row. Where the unweighted fit is consistent for the slope, the background is slightly shaded.

## C. Estimates with the controls based on child population taken in logarithms

### Table C–1. Reduced-form TWFE impact estimates for years of schooling, 1995

| | (1) | (2) | Hausman $p$: weighted vs. not |
|---|---|---|---|
| *With controls based on log number of children aged 5–14* | | | |
| Experiment | 0.14 | 0.10 | 0.26 |
| | (0.044) | (0.055) | |
| Placebo | 0.0096 | 0.072 | 0.13 |
| | (0.057) | (0.072) | |
| Kink | 0.094 | 0.072 | 0.68 |
| | (0.074) | (0.090) | |
| Data corrections | ✓ | ✓ | |
| Clustered | ✓ | ✓ | |
| Weighted | | ✓ | |
| Experiment $N$ | 78,470 | 78,470 | |
| Placebo $N$ | 78,488 | 78,488 | |
| Kink $N$ | 143,143 | 143,143 | |

*Notes:* Specifications are identical to those in panel A of Table 2, except that the child population controls are taken in logarithms.

### Table C–2. Reduced-form TWFE impact estimates for labor outcomes, 1995

| | Formal sector employment | | | Log hourly wage | | |
|---|---|---|---|---|---|---|
| | New | | | New | | |
| | (1) | (2) | Hausman $p$: weighted vs. not | (3) | (4) | Hausman $p$: weighted vs. not |
| *With controls based on log number of children aged 5–14* | | | | | | |
| Experiment | 0.026 | 0.024 | 0.61 | 0.033 | 0.036 | 0.76 |
| | (0.0047) | (0.0064) | | (0.012) | (0.016) | |
| Placebo | 0.0035 | 0.0079 | 0.40 | 0.0019 | –0.00063 | 0.84 |
| | (0.0061) | (0.0078) | | (0.015) | (0.019) | |
| Kink | 0.027 | 0.019 | 0.30 | 0.038 | 0.077 | 0.025 |
| | (0.0084) | (0.011) | | (0.020) | (0.027) | |
| Data corrections | ✓ | ✓ | | ✓ | ✓ | |
| Clustered | ✓ | ✓ | | ✓ | ✓ | |
| Weighted | | ✓ | | | ✓ | |
| Experiment $N$ | 78,470 | 78,470 | | 31,061 | 31,061 | |
| Placebo $N$ | 78,488 | 78,488 | | 30,458 | 30,458 | |
| Kink $N$ | 143,143 | 143,143 | | 57,234 | 57,234 | |

*Notes:* See notes to Table 2.

**Table C–3. 2SLS estimates of the effect of years of schooling on labor market outcomes, 1995**

| | Employment | | | | Log hourly wage | | | |
|---|---|---|---|---|---|---|---|---|
| | Instrument by young/old | | Kink instrument | | Instrument by young/old | | Kink instrument | |
| | (1) | (2) | (3) | (4) | (6) | (7) | (8) | (9) |
| *With controls based on log number of children aged 5–14* | | | | | | | | |
| Coefficient | 0.18 | 0.23 | 0.29 | 0.32 | 0.16 | 0.16 | 0.15 | 0.19 |
| | (0.062) | (0.13) | (0.22) | (0.37) | (0.053) | (0.071) | (0.081) | (0.084) |
| | | $[0.09, \infty)$ | $[0.08, \infty)$ | $(-\infty, -0.14] \cup [0.02, \infty)$ | $[0.06, 0.33]$ | $[0.01, 0.42]$ | $[-0.11, \infty)$ | $[0.05, 0.79]$ |
| KP $F$ | 10.1 | 3.44 | 1.78 | 0.78 | 11.2 | 8.56 | 3.62 | 6.02 |
| Data corrections | ✓ | ✓ | ✓ | ✓ | ✓ | ✓ | ✓ | ✓ |
| Clustered | ✓ | ✓ | ✓ | ✓ | ✓ | ✓ | ✓ | ✓ |
| Weighted | | ✓ | | ✓ | | ✓ | | ✓ |
| Pre-trend control | | | ✓ | ✓ | | | ✓ | ✓ |
| Observations | 382,383 | 382,383 | 606,781 | 606,781 | 113,738 | 113,738 | 169,020 | 169,020 |

*Notes:* See notes to Table 4 and Table 5.

# D. 2SLS estimates of the effect of years of schooling on labor outcomes, 1995, with instruments "by birth year"

**Table D–1. 2SLS estimates of the effect of years of schooling on labor outcomes, 1995, with instruments "by birth year"**

| | Employment | | | Log hourly wage | | |
|---|---|---|---|---|---|---|
| | (1) | (2) | (3) | (4) | (5) | (6) |
| *A. With controls based on number of children aged 5–14* | | | | | | |
| Coefficient | 0.101 | 0.10 | 0.090 | 0.0675 | 0.11 | 0.14 |
| | (0.0210) | (0.031) | (0.036) | (0.0280) | (0.048) | (0.052) |
| | | $(-\infty, \infty)$ | $(-\infty, \infty)$ | | $(-\infty, -0.24]$ | $(-\infty, -0.22] \cup$ $[0.38, \infty)$ |
| KP $F$ | | 1.86 | 1.04 | | 1.08 | 1.44 |
| Observations | | 152,989 | 152,989 | | 60,663 | 60,663 |
| *B. Without extra controls* | | | | | | |
| Coefficient | | 0.072 | 0.068 | | 0.046 | 0.12 |
| | | (0.030) | (0.039) | | (0.071) | (0.065) |
| | | $(-\infty, \infty)$ | $(-\infty, \infty)$ | | $(-\infty, -0.25]$ | $(-\infty, -0.14]$ |
| KP $F$ | | 1.07 | 0.66 | | 0.34 | 0.90 |
| Data corrections | | ✓ | ✓ | | ✓ | ✓ |
| Clustered | | ✓ | ✓ | | ✓ | ✓ |
| Weighted | | | ✓ | | | ✓ |
| Observations | 152,989 | 152,989 | 152,989 | 60,663 | 60,663 | 60,663 |

Notes: See notes to Table 4. The regressions reported here differ only in using the "by birth year" instrument set, which consists of interactions between $D_j$ and dummies for ages 2–24 as of 1974, except that age ≥12 defines one category. Columns 1 and 4 are copied from Duflo (2001), Table 7.

# E. Pooled results, all available years

The results presented below are computed identically to those in section 4 except that 1995 data is included.

**Table E–1. Reduced-form TWFE impact estimates for years of schooling, pooled data for 1995, 2005, 2011–2014, 2017–19**

| | (1) | (2) | Hausman $p$: weighted vs. not |
|---|---|---|---|
| *A. With controls based on number of children aged 5–14* | | | |
| Experiment | 0.11 | 0.098 | 0.68 |
| | (0.032) | (0.036) | |
| Placebo | 0.043 | 0.099 | 0.036 |
| | (0.037) | (0.042) | |
| Kink | 0.081 | 0.070 | 0.65 |
| | (0.037) | (0.040) | |
| *B. Without extra controls* | | | |
| Experiment | 0.081 | 0.079 | 0.93 |
| | (0.030) | (0.035) | |
| Placebo | 0.055 | 0.094 | 0.12 |
| | (0.035) | (0.040) | |
| Kink | 0.035 | 0.035 | 0.99 |
| | (0.035) | (0.038) | |
| Data corrections | ✓ | ✓ | |
| Clustered | ✓ | ✓ | |
| Weighted | | ✓ | |
| Experiment $N$ | 668,141 | 668,141 | |
| Placebo $N$ | 564,718 | 564,718 | |
| Kink $N$ | 1,166,869 | 1,166,869 | |

*Notes:* Notes to Table 2 apply. As dictated by availability of needed variables.

**Table E–2. Reduced-form TWFE impact estimates for labor outcomes, 1995, 2005, 2011–14**

| | Formal sector employment | | | Log hourly wage | | |
|---|---|---|---|---|---|---|
| | | | Hausman $p$: | | | Hausman $p$: |
| | (1) | (2) | weighted vs. not | (3) | (4) | weighted vs. not |
| *A. With controls based on number of children aged 5–14* | | | | | | |
| Experiment | 0.0057 | –0.00045 | 0.0018 | 0.021 | 0.0093 | 0.044 |
| | (0.0024) | (0.0031) | | (0.0073) | (0.0094) | |
| Placebo | –0.0035 | –0.018 | 0.00011 | –0.015 | –0.015 | 0.97 |
| | (0.0039) | (0.0057) | | (0.011) | (0.016) | |
| Kink | 0.010 | 0.014 | 0.28 | 0.025 | 0.035 | 0.35 |
| | (0.0048) | (0.0065) | | (0.012) | (0.018) | |
| *B. Without extra controls* | | | | | | |
| Experiment | 0.0011 | –0.0039 | 0.0082 | –0.0048 | –0.015 | 0.068 |
| | (0.0022) | (0.0028) | | (0.0070) | (0.0086) | |
| Placebo | –0.0035 | –0.015 | 0.0012 | –0.0092 | –0.012 | 0.75 |
| | (0.0036) | (0.0055) | | (0.0099) | (0.014) | |
| Kink | 0.0064 | 0.0080 | 0.64 | 0.010 | 0.015 | 0.59 |
| | (0.0045) | (0.0061) | | (0.011) | (0.017) | |
| Data corrections | ✓ | ✓ | | ✓ | ✓ | |
| Clustered | ✓ | ✓ | | ✓ | ✓ | |
| Weighted | | ✓ | | | ✓ | |
| Experiment $N$ | 460,853 | 460,853 | | 144,799 | 144,799 | |
| Placebo $N$ | 404,451 | 404,451 | | 99,396 | 99,396 | |
| Kink $N$ | 810,736 | 810,736 | | 241,957 | 241,957 | |

*Notes:* See notes to Table 3. The 2005 data covers employment but not wages.

**Table E–3. 2SLS estimates of the effect of years of schooling on labor market outcomes, pooled data for 1995, 2005, 2011–14**

| | Employment | | | | Log hourly wage | | | |
|---|---|---|---|---|---|---|---|---|
| | Instrument by young/old | | Kink instrument | | Instrument by young/old | | Kink instrument | |
| | (1) | (2) | (3) | (4) | (6) | (7) | (8) | (9) |
| *A. With controls based on number of children aged 5–14* | | | | | | | | |
| Coefficient | 0.059 | –0.0056 | 0.17 | 0.92 | 0.24 | 0.17 | –0.37 | –0.17 |
| | (0.026) | (0.039) | (0.14) | (3.05) | (0.085) | (0.16) | (0.74) | (0.24) |
| | [0.01, 0.17] | [−0.42, 0.11] | (−∞, −0.25] ∪ [−0.01, ∞) | (−∞, ∞) | [0.10, 1.09] | (−∞, ∞) | (−∞, 0.07] ∪ [0.19, ∞) | (−∞, 0.08] ∪ [0.40, ∞) |
| KP $F$ | 9.02 | 4.59 | 1.86 | 0.087 | 5.16 | 0.91 | 0.42 | 1.55 |
| *B. Without extra controls* | | | | | | | | |
| Coefficient | 0.016 | –0.055 | 1.35 | –0.29 | 0.59 | 0.32 | 0.013 | 0.010 |
| | (0.030) | (0.053) | (12.4) | (0.69) | (2.25) | (0.28) | (0.074) | (0.076) |
| | [−0.16, 0.12] | (−∞, 0.02] | (−∞, ∞) | (−∞, ∞) | (−∞, ∞) | (−∞, ∞) | (−∞, 0.15] | (−∞, 0.16] |
| KP $F$ | 5.29 | 4.03 | 0.012 | 0.25 | 0.048 | 0.82 | 4.60 | 5.08 |
| Data corrections | ✓ | ✓ | ✓ | ✓ | ✓ | ✓ | ✓ | ✓ |
| Clustered | ✓ | ✓ | ✓ | ✓ | ✓ | ✓ | ✓ | ✓ |
| Weighted | | ✓ | | ✓ | | ✓ | | ✓ |
| Pre-trend control | | | ✓ | ✓ | | | ✓ | ✓ |
| Observations | 460,853 | 460,853 | 858,355 | 858,355 | 144,358 | 144,358 | 249,612 | 249,612 |

**Figure E–1. Coefficients on interactions between age dummies and Inpres intensity in schooling regressions, 1995, 2005, 2011–14, 2017–19**

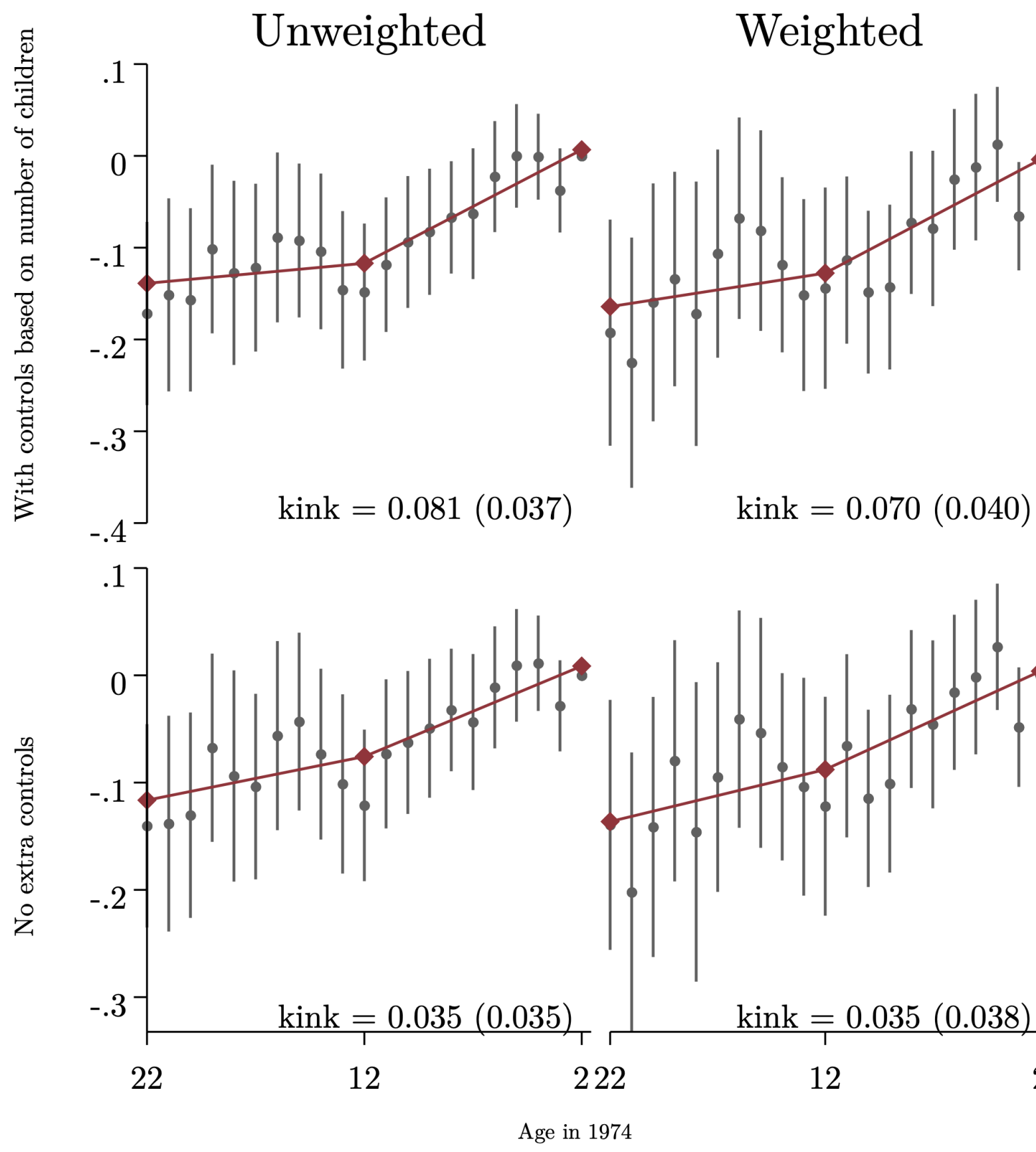

**Figure E–2. Coefficients on interactions between age dummies and Inpres intensity in labor outcome regressions, 1995, 2005, 2011–14**

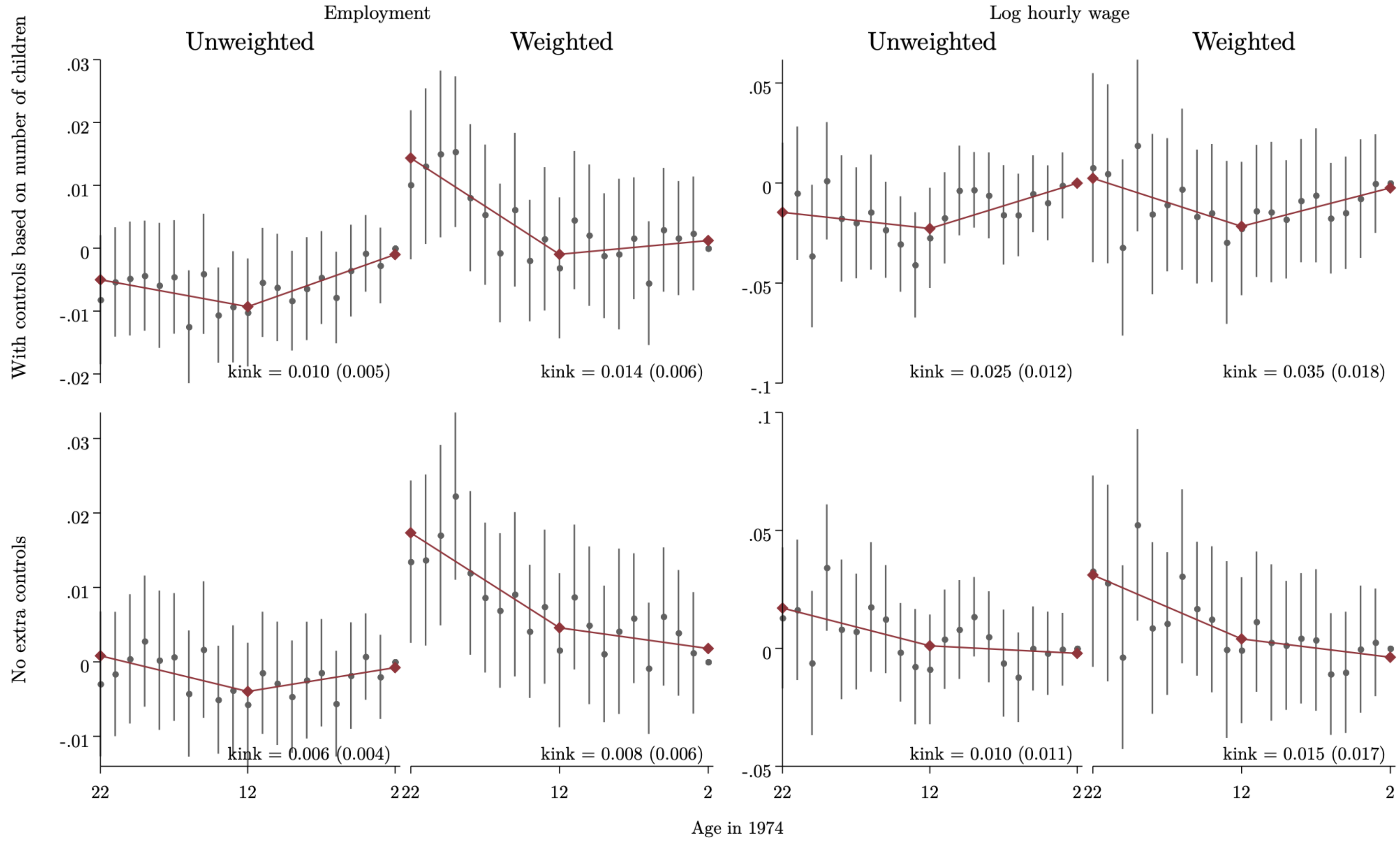

**Figure E–3. Confidence curves from wild-bootstrapped tests of the coefficient on schooling in 2SLS regressions in Table E–3**

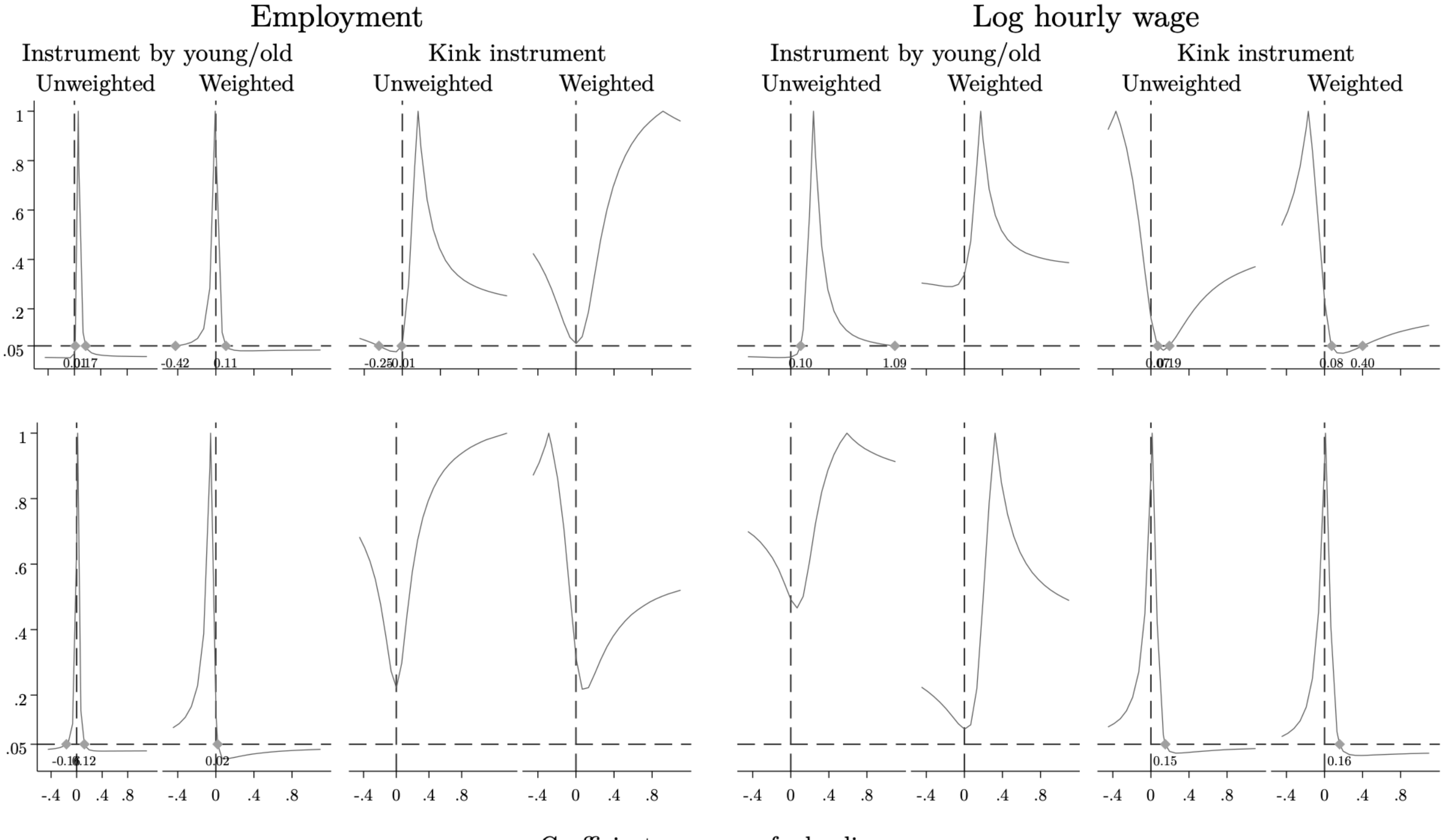

# F. Returns to primary school completion, 2011–14

**Table F–1. 2SLS estimates of the effect of years of primary completion on labor market outcomes, 2011–14**

| | Employment | | | | Log hourly wage | | | |
|---|---|---|---|---|---|---|---|---|
| | Instrument by young/old | | Kink instrument | | Instrument by young/old | | Kink instrument | |
| | (1) | (2) | (3) | (4) | (6) | (7) | (8) | (9) |
| *A. With controls based on number of children aged 5–14* | | | | | | | | |
| Coefficient | 0.11 | –0.14 | 0.73 | 0.84 | 1.41 | 0.34 | –6.48 | –3.03 |
| | (0.14) | (0.11) | (0.73) | (0.50) | (0.66) | (0.38) | (11.1) | (4.95) |
| | $[-0.19, 0.48]$ | $[-0.40, 0.07]$ | $(-\infty, \infty)$ | $[-0.00, \infty)$ | $[0.27, \infty)$ | $(-\infty, \infty)$ | $(-\infty, -0.26]$ | $(-\infty, \infty)$ |
| KP $F$ | 15.7 | 24.5 | 3.09 | 7.30 | 10.5 | 18.3 | 0.40 | 0.56 |
| *B. Without extra controls* | | | | | | | | |
| Coefficient | –0.21 | –0.32 | 0.45 | 0.42 | –1.02 | –1.15 | –0.85 | 1.04 |
| | (0.22) | (0.15) | (0.85) | (0.47) | (1.75) | (0.88) | (3.15) | (3.45) |
| | $(-\infty, 0.21]$ | $(-\infty, -0.08]$ | $(-\infty, \infty)$ | $(-\infty, \infty)$ | $(-\infty, \infty)$ | $(-\infty, 0.14]$ | $(-\infty, \infty)$ | $(-\infty, \infty)$ |
| KP $F$ | 7.25 | 15.3 | 1.68 | 6.06 | 1.62 | 6.62 | 0.51 | 0.41 |
| Data corrections | ✓ | ✓ | ✓ | ✓ | ✓ | ✓ | ✓ | ✓ |
| Clustered | ✓ | ✓ | ✓ | ✓ | ✓ | ✓ | ✓ | ✓ |
| Weighted | | ✓ | | ✓ | | ✓ | | ✓ |
| Pre-trend control | | | ✓ | ✓ | | | ✓ | ✓ |
| Observations | 382,383 | 382,383 | 705,366 | 705,366 | 113,297 | 113,297 | 188,949 | 188,949 |

**Figure F–1. Confidence curves from wild-bootstrapped tests of the coefficient on primary completion in 2SLS regressions in Table F–1**

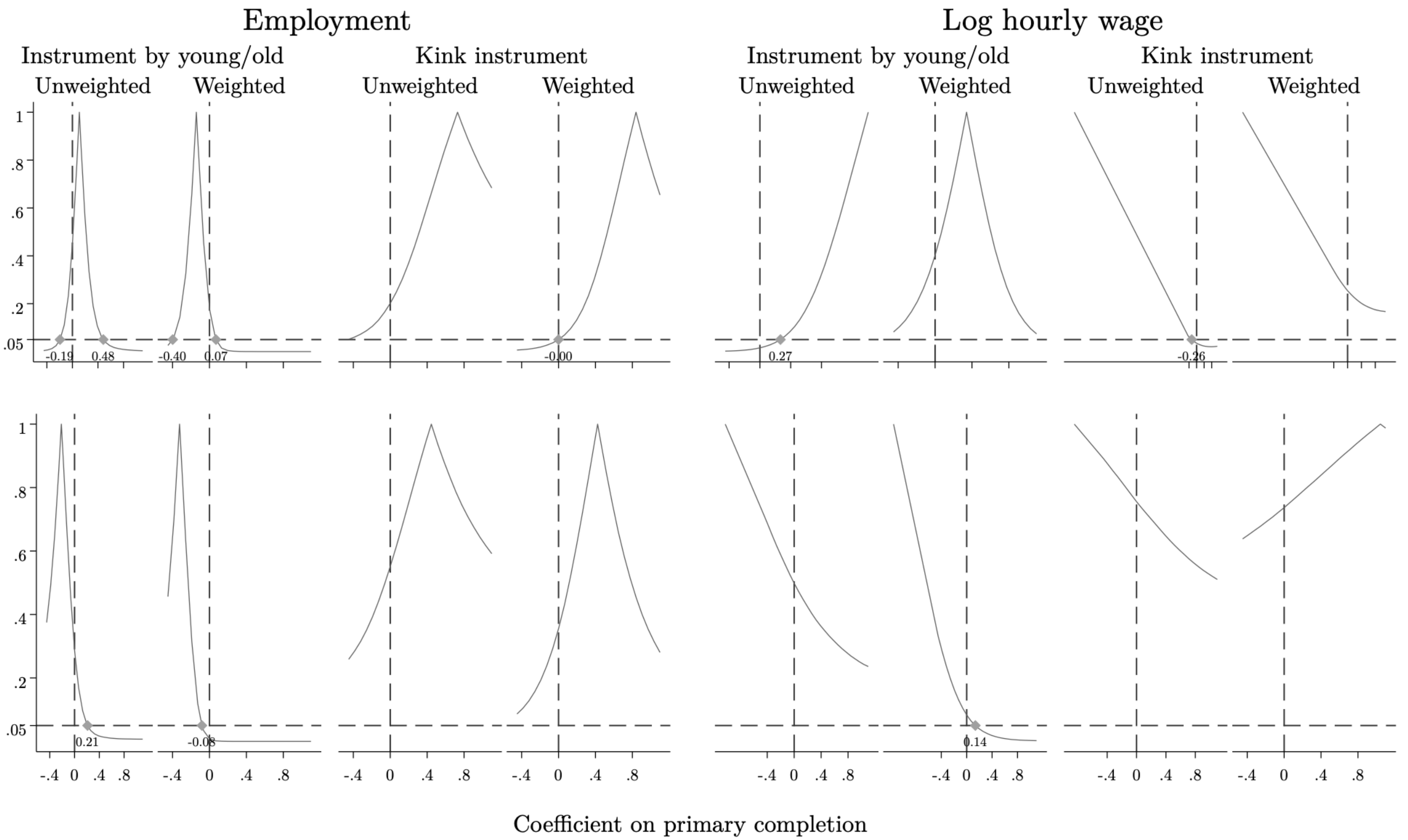

## G. Quantile-based estimation

The changes-in-changes (CIC) estimator of Athey and Imbens (2006) is defined for the 2×2 set-up. For a concrete example of its mechanics, suppose a person in a treatment group is observed before treatment to earn a wage of 1000 rupiah per hour. Suppose that wage rate would place the person in the 25$^{th}$ percentile of the *control* group's distribution for the same, pre-treatment period. Finally, suppose that the 25$^{th}$ percentile of the control group's post-treatment distribution is 2000 rupiah per hour. Then this treatment-group subject would contribute an observation of 2000 to the counterfactual distribution for the *treatment* group in the post-treatment period. The gap between a given quantile of this counterfactual distribution and the observed quantile is an estimate of the impact of treatment at that point in the distribution.

CIC is designed for continuous outcomes, which here restricts us to wages. (The estimator's results for schooling attainment, which is quantized by year, are degenerate.) Of necessity I apply CIC within a 2×2 set-up, with the same definitions of the high- and low- treatment groups and pre- and post-treatment cohorts as in the Duflo (2001) 2×2 DID. I incorporate the needed controls—dummies for year of birth, regency of birth, and year of survey—using the method of Melly and Santangelo (2015). I estimate in the 1995 sample, the 2011–14 sample, and the two together.

Estimates by decile appear in Figure G–1, along with bootstrap-based 95% confidence intervals. $p$ values from tests of various null hypotheses synopsizing the results are gathered in Table G–1. Especially when not weighting, the results for 1995 exhibit a dramatic swing toward negative impacts in the lower deciles. However, the Cramér–von Mises tests in the first two columns of in Table G–1 do not strongly reject the null of no impact at all deciles ($p = 0.30$ without weights and 0.59 with); perhaps the test is under-powered. At any rate, in the larger, pooled follow-up of 2011–14—in prime earnings years—there is a stronger suggestion of *positive* impact at lower deciles. The null of a positive impact at all deciles is rejected only at $p = 0.74$ when weighting, while the null of negative impact at all deciles is rejected at $p = 0.07$.

The finding of a benefit for poorer people is consistent with the story that Inpres increased schooling for the less-well-off, those for whom the marginal year of schooling was in primary school. The slight hint in of negative impacts at higher deciles might reflect the confounding from secondary school expansion discussed in section 5.

de Chaisemartin and D'Haultfœuille (CH, 2017) introduces a variant of CIC that offers a quantile-based way to estimate returns to schooling, as distinct from reduce-form estimation as just performed. Indeed, that paper illustrates its methods on the Duflo (2001) data, and estimates the wage return to a year of schooling at 0.099 log points (standard error 0.017). However, the CH Wald CIC estimator requires a control arm within which expected treatment—years of schooling—is stable over time, as well as arms in which expected treatment strictly increases or decreases. To accommodate this requirement, CH discards Duflo (2001)'s Inpres-based identification strategy. CH instead groups regencies by whether average schooling rose, fell, or stayed about the same between the age 12–17 and age 2–6 cohorts. In the instrumental variables perspective, this instruments schooling using a trichotomous factor variable that itself depends on schooling and thus is about as presumptively endogenous. As is demonstrated in appendix H, the combination of this procedure for forming supergroups and the Wald CIC estimator introduces endogeneity bias. I therefore do not view the CH results for schooling in Indonesia as much more informative as to causal impacts than OLS-type results.

**Table G–1. Rejection *p* values for various null hypotheses about the CIC results**

| Null | 1995 | | 2011–14 | | 1995, 2011–14 | |
|---|---|---|---|---|---|---|
| No effect, all deciles | 0.30 | 0.59 | 0.00 | 0.23 | 0.003 | 0.03 |
| Constant effect | 0.46 | 0.30 | 0.00 | 0.27 | 0.00 | 0.01 |
| Positive effect, all deciles | 0.24 | 0.26 | 0.56 | 0.74 | 0.43 | 0.55 |
| Negative effect, all deciles | 0.51 | 0.68 | 0.00 | 0.07 | 0.00 | 0.004 |
| Weighted | | ✓ | | ✓ | | ✓ |

Notes: All $p$ values are from Cramér–von Mises tests.

**Figure G–1. Changes-in-changes estimates of impact of high Inpres treatment on log hourly wage at various percentiles**

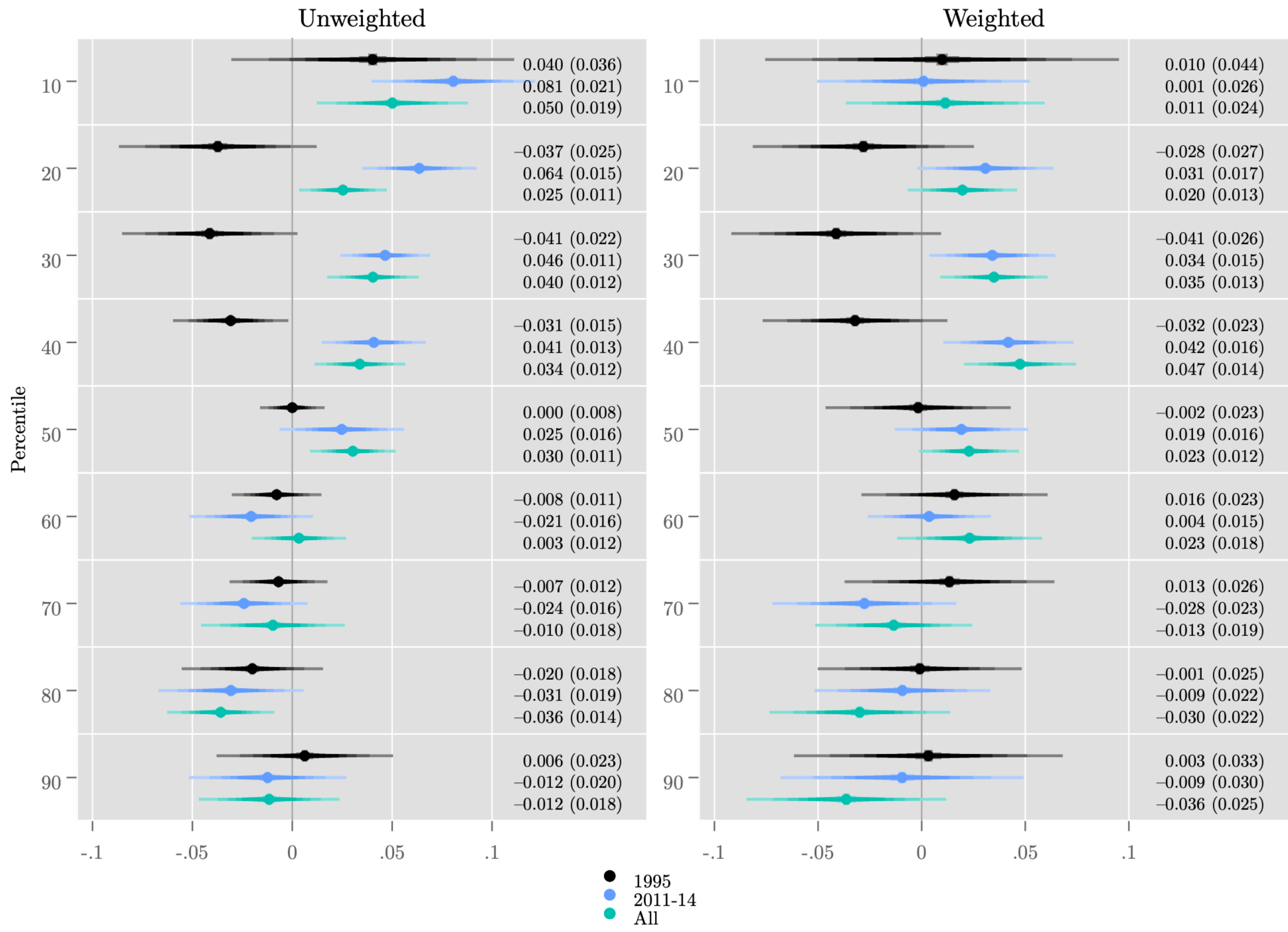

*Notes:* The plot shows, at various percentiles, estimates of the impact of high Inpres treatment on earnings, along with bootstrapped 95% confidence intervals. Point estimates and standard errors are printed on the right. Results are from the CIC-with-controls estimator of Melly and Santangelo (2015). The definitions of before and after periods and low- and high-treatment groups are the same as in the Duflo (2001) 2×2 DID regressions. The controls are dummies for regency of birth, year of birth, and year of survey.

## H. Simulation of de Chaisemartin and D'Haultfœuille (2017)'s Wald CIC

Appendix G asserts that a combination of procedures in de Chaisemartin and D'Haultfœuille (CH, 2017)— gathering groups into three supergroups, then performing "Wald CIC"—will produce biased impact estimates if treatment is endogenous. (In the CH application to Duflo (2001), "treatment" refers not to the basis of Duflo (2001)'s instruments, planned Inpres schools per child, but to schooling attainment, which is indeed presumed endogenous.) This appendix justifies the concern with a simulation.

The simulated data come from 100 groups, indexed by $j$, each with 100 subjects, indexed by $i$. Each subject is observed once, with equal probability in time period 0 or 1. The data generating process is

$$S_{ijt} = \lceil Z_{jt} + s_{jt} + u_{ijt} \rceil$$
$$Y_{ijt} = \beta S_{ijt} + v_{ijt}$$

$Z_{jt} \sim$ i.i.d. $\mathcal{N}(0,1)$ is an exogenous, standard-normal, group-level component of schooling. $s_{jt}$ is a group-level time trend such that $s_{j0} \equiv 0$ and $s_{j1} \sim$ i.i.d. $\mathcal{N}(0,1)$ for all $j$; it determines whether treatment in each group tends to rise, fall, or stay about the same. $u_{ijt}, v_{ijt} \sim \mathcal{N}(0,1)$ are error terms with cross-correlation $\rho$, which constitutes the endogeneity. Schooling, $S_{ijt}$, is discretized with the ceiling operator because Wald CIC is defined only when treatment has finite support. $Y_{ijt}$ is wages.

To collect the groups into supergroups characterized by downward, flat, or upward trends in treatment, CH suggest using a $\chi^2$ test with a high $p$ value, 0.5, to judge whether average treatment changes in each group between pre- and post-treatment periods. I apply that procedure to each simulated data set. Then I apply three estimators: 2SLS-based Wald DID on the original grouping, using the perfect instrument $Z$; Wald DID instrumenting with supergroup dummies; and CH's Wald CIC on the supergroups. The first estimator serves to benchmark the other two.

I run two sets of 100 simulations. In the first, $\rho = 0$, making the treatment $S_{ijt}$ exogenous.[31] In the second, $\rho = 0.95$, making treatment highly endogenous. In both, the impact parameter $\beta$ is 0.

Results from the simulations appear in Table H–1. They confirm that the Wald DID and CIC estimators run on the supergroups are unbiased only when treatment is exogenous. Then, the estimators produce nearly identical distributions centered on the correct value for $\beta$, zero. Introducing substantial endogeneity does not harm the Wald DID regressions with the perfect instrument, but lifts the mean estimates away from the true value of zero, by a standard deviation, when estimating from the CH supergroups.

[31] A few simulations fail when, in the Wald CIC, the support of the constructed counterfactual post-treatment outcome distribution for the treatment groups does not cover that for the observed outcome distribution, so that changes at one or more quantiles cannot be estimated.

**Table H–1. Average supergroup size and point estimates from DID-type estimators on simulated data**

| | $\rho = 0$ | | $\rho = 0.95$ | |
|---|---|---|---|---|
| | Mean | Standard deviation | Mean | Standard deviation |
| Number of groups in supergroups defined by trend in treatment | | | | |
| Decreasing | 46.9 | 5.1 | 46.6 | 5.2 |
| Flat | 7.5 | 2.5 | 8.1 | 5.4 |
| Increasing | 45.6 | 4.9 | 45.8 | 4.9 |
| | | | | |
| Estimate of $\beta$ | | | | |
| Wald DID on groups | 0.001 | 0.025 | 0.001 | 0.025 |
| Wald DID on supergroups | 0.001 | 0.021 | 0.020 | 0.020 |
| Wald CIC on supergroups | 0.001 | 0.022 | 0.018 | 0.023 |
| | | | | |
| Simulations | 98 | | 99 | |

*Notes:* True estimand in all simulations is 0. Simulated data sets consist of 100 groups with 100 subjects. Following de Chaisemartin and D'Haultfœuille (2017), groups are sorted into supergroups based on an $F$ test of whether their average schooling rises or falls, using a threshold $p$ value of 0.5. Wald DID is DID-type 2SLS on the original groups, instrumenting with the known, exogenous component of treatment. Wald DID and Wald CIC on supergroups are de Chaisemartin and D'Haultfœuille (2017) estimators and take the "flat" supergroup as the control group.